\documentclass[10pt,preprint,review]{elsarticle}
\usepackage{amsmath,amssymb,amsfonts,amsthm}
\usepackage{algorithmic}
\usepackage{array}
\usepackage{textcomp}
\usepackage{stfloats}
\usepackage{url}
\usepackage{verbatim}
\usepackage{graphicx}
\usepackage{subcaption}
\usepackage{bm}
\usepackage{float} 
\usepackage{booktabs}
\usepackage{array, caption, threeparttable}
\usepackage[font=small,labelfont=bf,labelsep=none]{caption}
\usepackage{balance}
\usepackage{algorithm}
\usepackage{natbib}
\setcitestyle{numbers,square}
\usepackage{color}
\usepackage{stfloats}
\usepackage{mathrsfs}
\newtheorem{Theorem}{Theorem}
\newtheorem{Proposition}[Theorem]{Proposition}

\makeatletter

\usepackage{hyperref}
\journal{}
\makeatother

\begin{document}
	
\begin{frontmatter}
	
\title{From Device to Dynamics: An Iterative Architectural Framework for High-Performance Single-Photon Detection at Room Temperature}

\author{Hao Shu\corref{cor1}}
\ead{Hao_B_Shu@163.com}
\cortext[cor1]{}

\affiliation{organization={Sun-Yat-Sen University},
	city={Shenzhen},
	country={China}}


\begin{abstract}
Photon detection is a cornerstone of quantum technology, traditionally regarded as a static device-level operation constrained by the intrinsic physical properties of single-photon detectors (SPDs). Consequently, high-performance detection has been heavily reliant on superconducting technologies, whose requirement for cryogenic temperatures imposes significant infrastructure burdens and limits scalable deployment. To circumvent these constraints, we propose the Enhanced Single-Photon Detection (ESPD) framework, which shifts the photon-detection paradigm from device-centric optimization to an integrated quantum-information-processing (QIP) task. By incorporating state preparation, controlled operations, projective measurements, and multi-copy decision analysis, we establish a nonlinear dynamical model that reformulates detection as an iteratively enhanced process. This architecture enables systematic performance upgrades through structural design rather than material modification, allowing high-performance detection with exclusively room-temperature hardware. Through analytical approximations, Monte Carlo analysis, and numerical simulations, grounded in parameters derived from non-superconducting components, we show that the ESPD dynamics converge to a high-performance basin of attraction even when initialized by low-performance SPDs. Specifically, the framework can upgrade a conventional room-temperature SPD to achieve an effective detection efficiency (DE) exceeding 93\% with a dark count rate (DCR) below $10^{-9}$, metrics comparable to state-of-the-art superconducting nanowire SPDs (SNSPDs). Such enhancements significantly lower the threshold for the required minimum channel transmission rate in quantum communication applications. While physical realization requires further component integration efforts, this work establishes a rigorous theoretical foundation for enhancing detection via architectural QIP principles. It provides not only a blueprint for next-generation room-temperature photon detection but also a general methodology for transcending device-level constraints in broader quantum technologies.
\end{abstract}


\end{frontmatter}

\section{Introduction}
\label{Int}

Photon detection is a foundational pillar of quantum optics, providing essential interfaces for nearly all quantum information applications. Traditionally, the detection process has been conceptualized as a static, device-level operation, whose performance is regarded as intrinsic to the single-photon detector (SPD) material properties. The utility of an SPD is primarily governed by two competing metrics: detection efficiency (DE, denoted by $\eta$), which represents the probability of detecting a non-vacuum incident photon, and the dark count rate (DCR, denoted by $d$), which represents the probability of returning an incorrect positive detection report in the absence of a photon.

High DE is indispensable for obtaining meaningful outcomes in quantum experiments. In fundamental tests of quantum nonlocality, verifying violations of the Clauser-Horne-Shimony-Holt or Eberhard inequalities requires DE exceeding strict thresholds to close detection loopholes~\cite{GG1999A,MB2023Bounding,QA2012Maximal,VP2010Closing,RK2001Experimental,M2023Efficiency,WR2012Loophole}, such as approximately 67\%~\cite{E1993Background} or 83\%~\cite{CH1974Experimental}. In optical quantum computation, where successful events can be probabilistic and require the simultaneous detection of multiple photons, the overall success probability decreases exponentially with imperfect DE~\cite{KL2001scheme,DG2025High,SR2005Thresholds}. In heralded single-photon sources, where one photon of an entangled pair signals the presence of the other, the achievable heralding rate is also strictly limited by DE~\cite{RM2013Highly,NA2015Ultra,DM2022Improved,SC2023High}.

Regarding DCR, a particularly stringent constraint arises in quantum communication, notably quantum key distribution (QKD). The security of practical QKD protocols relies on maintaining the quantum bit error rate (QBER) below protocol-dependent thresholds that are usually lower than 11\%~\cite{BB1984Quantum,GL2004Security,LY2018Overcoming,WH2019Beating,ZZ2022Mode}. However, optical communication channels inevitably induce exponential photon attenuation with distance~\cite{KY1986Transmission,NK2002Ultra,HH2013Record,TS2017Lowest}. When the effective signal rate, determined by exponentially decay channel transmission for a given DE, approaches the DCR floor of the SPDs, detection events become dominated by dark counts that contribute a 50\% error rate, causing the QBER to rise precipitously. Consequently, the achievable secure distance of QKD systems is ultimately bounded by the DCR of the employed SPDs~\cite{GL2004Security,S2023Solve,S2025Elimiating}.

Motivated by these critical demands, the development of high-performance SPDs has remained a primary research focus. While conventional semiconductor-based SPDs offer the advantage of non-cryogenic operation, they typically suffer from intrinsic limits on relatively low DE and high DCR. For instance, a frequency upconversion SPD reported in 2004 operates at 300~K with a DE of approximately 59\% and a DCR of $10^{-2}$~\cite{AW2004Efficient}, while an InGaAs/InP SPD reported in 2017 operates at 223~K achieved a DE of about 27.5\% with a DCR of $10^{-6}$~\cite{JL20171.25}. Such specifications often fall short of advanced quantum task requirements, which frequently demand DE exceeding 90\% and DCR below $10^{-7}$.

To meet these stringent requirements, substantial efforts have gravitated toward superconducting SPD technologies, most notably superconducting nanowire SPDs (SNSPDs)~\cite{HL2020Detecting,CL2021Detecting,XZ2021Superconducting,CK2023High}. These devices have successfully pushed DE beyond 90\% while suppressing DCR by several orders of magnitude compared to semiconductor counterparts, thereby enabling demanding applications such as long-distance QKD~\cite{WH2019Beating,CZ2020Sending,ZL2023Experimental,LZ2023Experimental}. Recent representative benchmarks of state-of-the-art superconducting SPDs are summarized in Table~\ref{SPD}.

\begin{table}[htbp]
\caption{\textbf{Recent representative benchmarks of superconducting SPDs.} OT: Operating Temperature; DE: Detection Efficiency; DCR: Dark Count Rate}
\label{SPD}
\centering
\begin{tabular}{ccccc}
\hline
Year & Researchers (et al.) & OT & DE & DCR\\
\hline
2020 & P. Hu\qquad\ \cite{HL2020Detecting} & 2.10 K & 95.0 \% & $0.5\times 10^{-5}$\\
2021 & J. Chang\quad\cite{CL2021Detecting} & 2.50 K & 99.5 \% & $1.1\times 10^{-3}$\\
2021 & G. Z. Xu\quad\cite{XZ2021Superconducting} & 0.84 K & 92.2 \% & $3.6\times 10^{-5}$ \\
2023 & I. Craiciu\quad\cite{CK2023High} & 0.90 K & 78.0 \% & $1.0\times 10^{-7}$\\
2024 & I. Charaev\ \cite{CB2024Single} & 20.0 K & 7.60 \% & $1.0\times 10^{-3}$\\
\hline
\end{tabular}
\end{table}

However, the superior performance of superconducting SPDs comes at the cost of cryogenic operation. This imposes stringent environmental requirements and necessitates bulky, power-intensive cooling infrastructure. The significant degradation of performance at elevated temperatures highlights a strong trade-off between detection performance and operating conditions. These constraints present major barriers to scalable deployment, particularly in field applications, satellite payloads, and scenarios with strict size, weight, and power limitations where cryogenic infrastructure is impractical.

To address the limitations of room-temperature SPDs without resorting to cryogenic conditions, we introduce the enhanced single-photon detection (ESPD) framework, which shifts the focus from device-centric optimization to a system-level architectural paradigm with nonlinear dynamics. Within this framework, photon detection is reformulated as a composite quantum-information-processing task that integrates state preparation, controlled quantum operations, projective measurements, and multi-copy decision analysis, rather than relying solely on SPD material properties. By formulating this multi-stage enhancement approach as a non-linear dynamical system, we analyze the convergence range and steady-state performance of the ESPD structure, providing a novel perspective on hardware noise suppression through iterative convergence processing. Our analysis demonstrates that this framework can elevate the performance of conventional, room-temperature SPDs to levels comparable to, or even exceeding, those of cryogenic superconducting counterparts.

Numerical simulations, grounded in physically motivated parameters from commercially available devices, indicate that the iterative mapping within the ESPD framework exhibits strong convergence toward high-performance regimes. Monte Carlo analysis reveals that, provided the initial SPD parameters fall within a broad basin of attraction (i.e., not starting with an extremely poor SPD), the evolved DE can surpass 90\% alongside ultra-low DCR. For instance, starting from a legacy SPD (e.g., DE $\approx 59\%$ and DCR $\approx 10^{-2}$~\cite{AW2004Efficient}), the system converges to an effective DE exceeding 93\% and an effective DCR below $10^{-9}$. Such performance rivals or even outperforms state-of-the-art superconducting SPDs, substantially relaxing the minimal tolerable channel transmission rate in QKD, and is achieved under room-temperature settings, circumventing the need for cryogenic infrastructure.

While physical implementation will require further system-level integration of quantum technologies, the ESPD framework provides a structured nonlinear dynamical formulation for room-temperature high-performance photon detection. Furthermore, we offer a practical roadmap to guide future experimental efforts by explicitly discussing technological feasibility and analyzing the trade-offs between resource overhead and operating temperature. Ultimately, the design principles and perspective shift offered by the ESPD framework serve as a theoretical blueprint for overcoming intrinsic device limitations through dynamical process design, which can boost the development of a broader range of quantum devices beyond photon detection and unlock diverse high-performance quantum applications at room temperature.

\section{Methods and Results}
\label{Results}

This section details the ESPD architecture and formulates the corresponding detection dynamics. The ESPD paradigm is introduced in Section \ref{sec:Design}, with rigorous theoretical formulations established in Section~\ref{sec:Theory}. Analytical approximations are developed in Section \ref{sec:Approximation}, followed by (Monte Carlo) stability analysis and numerical simulations in Section \ref{sec:Experiments}. Finally, the implications for QKD applications are discussed in Section \ref{sec:Application}.

\subsection{Paradigm of the Enhanced Single-Photon Detection}
\label{sec:Design}

\begin{figure}[htbp]
    \centering
    \includegraphics[width=\linewidth]{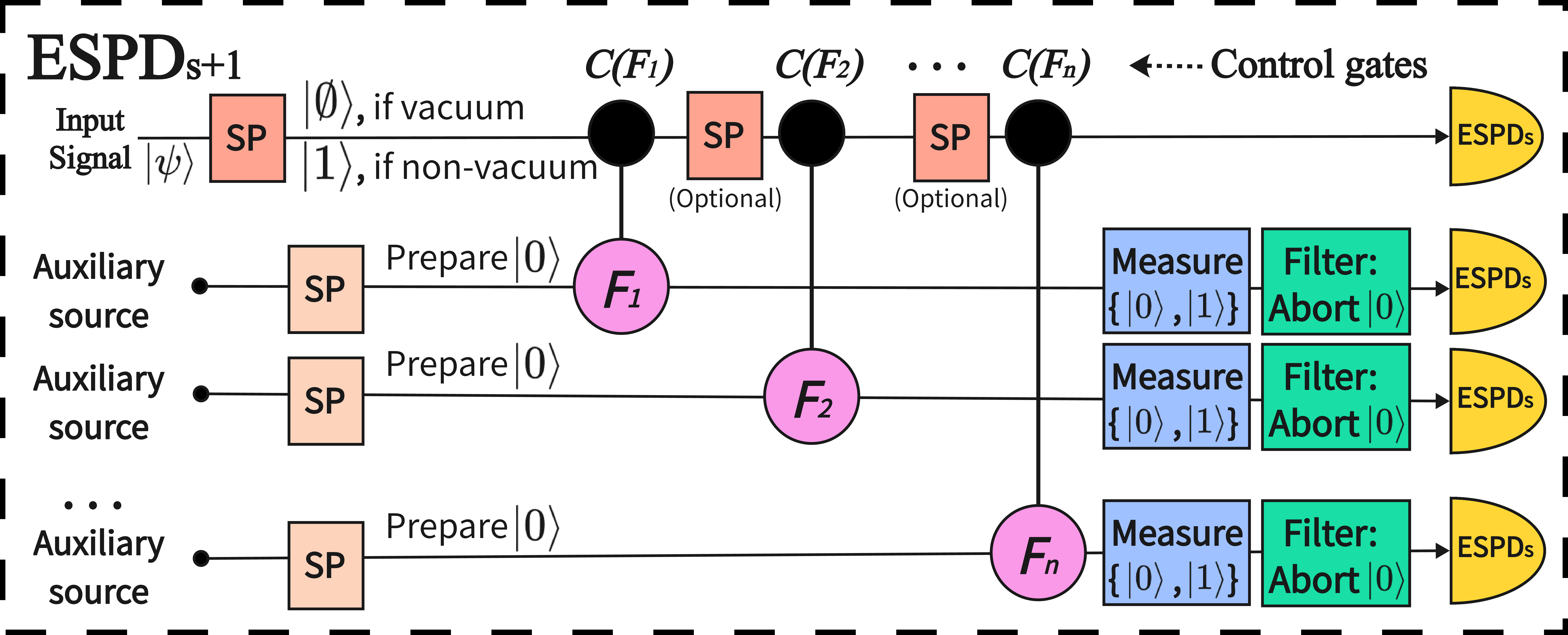}
    \caption{\ Schematic representation of recursive ESPD architecture. The $(s+1)$-th level detector ($\text{ESPD}_{s+1}$) is constructed by integrating multiple copies of the $s$-th level detector ($\text{ESPD}_{s}$) with state preparation, controlled gates, and projective measurements. Here, $C(F_{i})$ are controlled gates, SP are state-preparation operations, measurements are projective in the basis $\{|0\rangle,|1\rangle\}$, and filters indicate the abortion of the $|0\rangle$ path.}
    \label{fig:Design}
\end{figure}

The ESPD architecture is formulated as a recursive, level-by-level framework in which baseline SPD performance is systematically enhanced through successive processing stages. Within this paradigm, the performance is characterized by the effective DE ($\eta_{s}$) and DCR ($d_{s}$) at level $s$. Recall that DE is the probability of generating a positive detection event given a non-vacuum (single-photon) input, and DCR is the probability of generating an (incorrect) positive report given a vacuum input. Let $\text{ESPD}_{s}$ denote the established detector at level $s$, characterized by $(\eta_{s}, d_{s})$. The evolution begins with an existing SPD, denoted as $\text{ESPD}_{0}$, with initial values $\eta_{0}=\eta$ and $d_{0}=d$, the performance of the physical baseline SPD. The schematic of this recursive logic is illustrated in Fig.~\ref{fig:Design}.

The construction of the $(s+1)$-th level detector, $\text{ESPD}_{s+1}$, is defined as follows. As a representative example, an implementation can be considered with the degree of freedom (DOF) using photon polarization and the controlled operations using controlled-NOT (C-NOT) gates.

\begin{enumerate}
    \item \textbf{DOF Selection:} 
    Select a manipulable DOF of the incoming photon (for example, polarization). The computational basis for this DOF is denoted by $\{|0\rangle,|1\rangle\}$. Hence, the signal is modeled as a qutrit in the basis $\{|\emptyset\rangle, |0\rangle, |1\rangle\}$, where $|\emptyset\rangle$ denotes the vacuum state, and $|0\rangle$ and $|1\rangle$ denote orthonormal states of a non-vacuum photon\footnote{If the selected DOF is the photon number itself, $|0\rangle$ coincides with $|\emptyset\rangle$ and $|1\rangle$ represents a non-vacuum state, reducing the description to qubit. In practical architectures, an internal DOF such as polarization, distinct from photon existence, is often more convenient.}.
    
    \item \textbf{State Preparation:} 
    Apply the state-preparation operation, initializing the selected DOF to $|1\rangle$. As a result, a non-vacuum signal is mapped to $|1\rangle$, while a vacuum input remains in state $|\emptyset\rangle$.
    
    \item \textbf{Controlled Operations:} 
    Operate a set of $n_{s+1}$ controlled operations, denoted as $C(F_{i})$ for $i=1,\dots,n_{s+1}$ (typically C-NOT gates), coupling the signal path with $n_{s+1}$ auxiliary photons, each initialized in state $|0\rangle$.
    
    \item \textbf{Auxiliary Measurement:} 
    Perform projective measurements on the auxiliary outputs in basis $\{|0\rangle, |1\rangle\}$. Outcome paths corresponding to $|0\rangle$ are discarded; only the $|1\rangle$ outcomes are retained for subsequent detections.
    
    \item \textbf{Detection:} 
     Carried out detections on all remaining paths, including both the signal path and the auxiliary paths, using $s$-th level detectors ($\text{ESPD}_{s}$).
    
    \item \textbf{Decision Logic:} 
    A positive detection report for $\text{ESPD}_{s+1}$ is generated if and only if at least $k_{s+1}$ positive detection events occur among the $n_{s+1}+1$ detectors, consisting of $n_{s+1}$ auxiliary detectors and one detector on the signal path. Otherwise, a negative report is returned. Here, $k_{s+1}$ is a preselected threshold integer, and $n_{s+1}$ denotes the number of effective controlled operations\footnote{The main discussions of the article assume deterministic controlled gates, for which $n_{s+1}$ equals the number of implemented operations. For probabilistic gates, $n_{s+1}$ refers to the number of successful operations after post-selection. In this case, the decision logic is naturally reformulated in terms of the fraction of positive detection outcomes among the effective detections exceeding a predefined threshold.}.
\end{enumerate}

\textbf{Physical intuition:} From a functional perspective, each level of the ESPD framework increases the number of effective detection opportunities available to an input signal. In an idealized limit where all components except the SPDs are assumed to be perfect, the $(s+1)$-th level produces $n_{s+1}$ auxiliary signal instances identical to the input one through the controlled modules, which are independently detected by the $n_{s+1}$ ESPDs. A non-vacuum outcome is reported if at least $k_{s+1}$ positive detection events are registered among the $n_{s+1}+1$ detections. Therefore, the effective DE increases with decreasing threshold $k_{s+1}$; for instance, when $k_{s+1}=1$, the effective DE is given by $1-(1-\eta_{s})^{n_{s+1}}$. Furthermore, the effective DCR is suppressed because a false positive outcome requires at least $k_{s+1}$ independent dark count events, leading to an approximate probability of $d_{s}^{k_{s+1}}$. This architectural logic transforms detection from a single-shot device event into a more robust collective process, where multi-copy analysis can suppress the errors introduced by SPD imperfections. This intuition extends beyond the idealized limit when imperfections in the auxiliary components introduce errors that remain small compared to the activation probability of a genuine input signal. A rigorous treatment of these dynamics, accounting for non-ideal devices and error propagation, follows in Section~\ref{sec:Theory}.

\subsection{Theoretical Formulations} 
\label{sec:Theory} 
This subsection develops the rigorous theoretical foundation of the ESPD framework. We first define the governing parameters and underlying assumptions, followed by deriving the analytical expressions of the ESPD dynamics.

\subsubsection{Notations and Basic Assumptions}

The evolution of the system is governed by a set of parameters characterizing both detection components and coupling modules.

As defined previously, $\eta_{s}$ and $d_{s}$ represent the effective DE and DCR of the $\text{ESPD}_{s}$ unit, respectively. For analytical tractability, $\eta_{s}$ is treated as a unified value under a given photon-number distribution in the system. The architectural configuration at each level is determined by the number of effective controlled operations $n_{s}$ and the decision threshold $k_{s}$.

We assume that all controlled operations $C(F_{i})$ are identical (denoted as $C(F)$) and implemented in a feed-forward configuration, such that $n_{s}$ represents both the number of effective modules and the total number of physical modules required at level $s$. However, the generalized ESPD paradigm remains valid for post-selection controlled gates (see Section~\ref{sec:ControlledGate} for a brief discussion). Let $p$ be the transmission probability of the signal path through a single module (an (optional) state preparation followed by a $C(F)$), accounting for incidental losses. Specifically, $p$ is the probability that the output signal remains non-vacuum given a non-vacuum input of a single non-cascaded module, accounting for loss but irrespective of decoherence (since the state can be re-prepared). We further define $P$ and $Q$ as the conditional probabilities of obtaining a non-vacuum output on the auxiliary (controlled) path before detection, given a non-vacuum or vacuum input to the module, respectively. Thus, $Q$ encapsulates the intrinsic error rate of the auxiliary path attributable to state preparation and measurement (SPAM), not including the final detection, in the absence of $C(F)$. 

It is crucial to note that $p$, $P$, and $Q$ represent the intrinsic performance metrics of a single, non-cascaded controlled module. These parameters are defined at the component level and remain invariant with respect to the number of cascaded modules. Their cumulative impact on the system’s evolutionary trajectory is fully captured by the recursive relations derived in Section \ref{sec:Formulation}.

\subsubsection{Parameter Approximation Assumptions}
\label{sec:Assumptions}

While the general formulation remains exact, the following assumptions, which are physically and practically motivated, are introduced to facilitate the analytical approximations in Section~\ref{sec:Approximation}. Such assumptions are not required for the general theoretical formulation of the ESPD framework nor for the numerical simulations and will be discussed in detail in Section~\ref{sec:Device}.

In practical scenarios, SPD DE typically ranges from 10\% to 90\%, while the DCR is generally below $10^{-3}$. Consequently, the hierarchy $0 \leq d \ll \eta \leq 1$ holds. It is further assumed that $d_{s} \ll P\eta_{s}$: If this condition were violated, the signal contribution from the controlled operations would become indistinguishable from device noise, rendering the operations practically insignificant. Additionally, the transmission rate of a single controlled module is assumed to satisfy $p \approx 1$, and the SPAM error of the auxiliary path is assumed to satisfy $Q \ll 1$, consistent with the maturity of commercial optical components (see Section~\ref{sec:Device}). Finally, considering practical implementation costs, $n_s$ and $k_s$ are taken to be moderate integers satisfying $1 \leq k_{s} \leq n_{s}$.

\subsubsection{Evolutionary Equation Formulations}
\label{sec:Formulation}

To describe the state transition from level $s$ to $s+1$, we first evaluate the reporting probabilities on the auxiliary and signal paths. Let $P_{s+1}$ and $Q_{s+1}$ be the probabilities of obtaining a positive report on an auxiliary path using $\text{ESPD}_{s}$ as the detector, given a non-vacuum and vacuum input to the $C(F)$ module, respectively:
\begin{equation}\label{P_{s+1}}\begin{aligned}
P_{s+1}=P[\eta_{s}+(1-\eta_{s})d_{s}]+(1-P)d_{s}=P\eta_{s}(1-d_{s})+d_{s}\approx P\eta_{s}
\end{aligned}\end{equation}
\begin{equation}\label{Q_{s+1}}\begin{aligned}
Q_{s+1}&=Q[\eta_{s}+(1-\eta_{s})d_{s}]+(1-Q)d_{s}=Q\eta_{s}(1-d_{s})+d_{s}\approx Q\eta_{s}+d_{s}\leq Q+d_{s}
\end{aligned}\end{equation}
Similarly, let $P'_{s}$ and $Q'_{s}$ denote the probabilities of obtaining a positive report on the signal path after all $C(F)$ operations, given a non-vacuum and vacuum output from the final controlled gate, respectively, then:
\begin{equation}\label{P'_{s+1}}\begin{aligned}
P'_{s+1}=\eta_{s}+(1-\eta_{s})d_{s}\approx \eta_{s},\quad Q'_{s+1}=d_{s}
\end{aligned}\end{equation}

The performance of $\text{ESPD}_{s+1}$ is determined by the collective response of $n_{s+1}+1$ detectors. For the DE calculation, a non-vacuum input is assumed, and the probability of obtaining at least $k_{s+1}$ positive reports out of $n_{s+1}+1$ total detections is evaluated. If the photon is lost exactly after the $i$-th $C(F)$, the probability of a final positive report is:
\begin{equation}\label{DEloss}\begin{aligned}
&P_{s+1,k_{s+1},i}\\
=&(1-Q'_{s+1})\sum_{j_{1}+j_{2}= k_{s+1}}^{n_{s+1}}\binom{i}{j_{1}}P_{s+1}^{j_{1}}(1-P_{s+1})^{i-j_{1}}\binom{n-i}{j_{2}}Q_{s+1}^{j_{2}}(1-Q_{s+1})^{n-i-j_{2}}
\\
&+Q'_{s+1}\sum_{j_{1}+j_{2}= k_{s+1}-1}^{n_{s+1}}\binom{i}{j_{1}}P_{s+1}^{j_{1}}(1-P_{s+1})^{i-j_{1}}\binom{n-i}{j_{2}}Q_{s+1}^{j_{2}}(1-Q_{s+1})^{n-i-j_{2}}
\end{aligned}\end{equation}
where the first and second terms correspond to scenarios that the signal path detector reports negative and positive, respectively.

If the photon survives all $C(F)$ operations, the probability of a final positive report is:
\begin{equation}\label{DElossless}\begin{aligned}
P_{s+1,k_{s+1}}=&P'_{s+1}\sum_{j\geq k_{s+1}-1}\binom{n_{s+1}}{j}P_{s+1}^{j}(1-P_{s+1})^{n_{s+1}-j}
\\
&\qquad\qquad\qquad\ +(1-P'_{s+1})\sum_{j\geq k_{s+1}}\binom{n_{s+1}}{j}P_{s+1}^{j}(1-P_{s+1})^{n_{s+1}-j}
\\
=&P'_{s+1}\binom{n_{s+1}}{k_{s+1}-1}P_{s+1}^{k-1}(1-P_{s+1})^{n_{s+1}-k_{s+1}+1}
\\
&\qquad\qquad\qquad\qquad\qquad\quad+\sum_{j\geq k_{s+1}}\binom{n_{s+1}}{j}P_{s+1}^{j}(1-P_{s+1})^{n_{s+1}-j}
\end{aligned}\end{equation}

The total effective DE at level $s+1$ is then the weighted sum:
\begin{equation}\label{DE}\begin{aligned}
\eta_{s+1}=p^{n_{s+1}}P_{s+1,k_{s+1}}+\sum_{i=1}^{n_{s+1}}p^{i-1}(1-p)P_{s+1,k_{s+1},i}
\end{aligned}\end{equation}

For the DCR calculation, a vacuum input is assumed. The probability of a false positive is given by:
\begin{equation}\label{DCR}\begin{aligned}
d_{s+1}=(1-Q'_{s+1})\sum_{j\geq k_{s+1}}&\binom{n_{s+1}}{j}Q_{s+1}^{j}(1-Q_{s+1})^{n_{s+1}-j}
\\
&+Q'_{s+1}\sum_{j\geq k_{s+1}-1}\binom{n_{s+1}}{j}Q_{s+1}^{j}(1-Q_{s+1})^{n_{s+1}-j}
\end{aligned}\end{equation}

In summary, the transition of performance metrics follows a non-linear dynamical system:
\begin{equation}\label{DynamicSystem}\begin{aligned}
(\eta_{s+1},d_{s+1})=G_{s+1}(\eta_{s},d_{s}),\quad (\eta_{0},d_{0})=(\eta,d)
\end{aligned}\end{equation}
where the map $G_{s+1}$ is defined by the recursive operators in Eqs.~\eqref{P_{s+1}} to \eqref{DCR}. 

The ESPD architecture thus induces a nonlinear dynamics whose fixed points, stability, and basins of attraction determine the asymptotic detection performance. Particularly, the fixed point $(\eta_{\infty}, d_{\infty})$ represents the fundamental physical limit of the ESPD framework under given architectural parameters, while the basins of attraction define the range of hardware-level imperfections that can be effectively mitigated to achieve such enhancement.

\subsection{Analytical Approximations Analysis}
\label{sec:Approximation}

While Eq.~\eqref{DynamicSystem} describes the exact dynamics of the ESPD framework, an explicit analytical solution for the iterated map is generally intractable. However, by employing the approximations in Eqs.~\eqref{P_{s+1}} to \eqref{P'_{s+1}} alongside the physically-motivated assumptions in Section~\ref{sec:Assumptions}, we can derive simplified bounds to elucidate the system's evolutionary behavior.

 Consider the evolution of DCR first. Eq.~\eqref{DCR} can be reformulated as:
\begin{equation}\label{RDCR}\begin{aligned}
d_{s+1}=Q'_{s+1}\binom{n_{s+1}}{k_{s+1}-1}Q_{s+1}^{k-1}(1-&Q_{s+1})^{n_{s+1}-k_{s+1}+1}
\\
&+\sum_{j\geq k_{s+1}}\binom{n_{s+1}}{j}Q_{s+1}^{j}(1-Q_{s+1})^{n_{s+1}-j}
\end{aligned}\end{equation}

To analyze the monotonicity of this transition, we introduce the following proposition.
\begin{Proposition}
    Define the function
    \begin{equation}\begin{aligned}
f(x)=a\binom{n}{k-1}x^{k-1}(1-x)^{n-k+1}+\sum_{j\geq k}\binom{n}{j}x^{j}(1-x)^{n-j}
\end{aligned}\end{equation}
for $a\geq 0$, then $f(x)$ is monotonically increasing for $x\in[0,\frac{k-1}{n}]$.
\end{Proposition}
\textbf{Proof:} For $x\in[0,\frac{k-1}{n}]$: 
\begin{equation}\begin{aligned}
 &\frac{d}{dx}f(x)
 \\
 =&x^{k-2}[\binom{n}{k-1}a(1-x)^{n-k}[(k-1)(1-x)-x(n-k+1)]
 \\
 &\qquad\qquad\qquad\quad+\sum_{j\geq k}\binom{n}{j}[jx^{j+1-k}(1-x)^{n-j}-x^{j-k+2}(1-x)^{n-j-1}(n-j)]]
 \\
 =&x^{k-2}[\binom{n}{k-1}a(1-x)^{n-k}(k-1-nx)+\sum_{j\geq k}\binom{n}{j}x^{j+1-k}(1-x)^{n-j-1}(j-nx)]
 \\
 \geq&x^{k-2}[\binom{n}{k-1}a(1-x)^{n-k}+\sum_{j\geq k}\binom{n}{j}x^{j+1-k}(1-x)^{n-j-1}](k-1-nx)\geq 0
\end{aligned}\end{equation}
$\blacksquare$

By applying the upper bound from Eq.~\eqref{Q_{s+1}}, substituting Eq.~\eqref{P'_{s+1}} into Eq.~\eqref{RDCR}, and assuming $k-1\geq nx$, we obtain the following bound for the DCR evolution:
\begin{equation}\label{IDCR}\begin{aligned}
d_{s+1}\leq d_{s}\binom{n_{s+1}}{k_{s+1}-1}(Q+d_{s})&^{k_{s+1}-1}(1-Q-d_{s})^{n_{s+1}-k_{s+1}+1}
\\
&+\sum_{j\geq k_{s+1}}\binom{n_{s+1}}{j}(Q+d_{s})^{j}(1-Q-d_{s})^{n_{s+1}-j}
\end{aligned}\end{equation}

Notably, this bound is independent of the DE $\eta_{s}$. Given $Q, d_{s} \ll 1$ and moderate values for the architectural parameters $(n, k)$, we can assume $\binom{n_{s+1}}{j}(Q+d_{s}) \ll 1$. This yields the approximation:
\begin{equation}\label{IDCR_approx}\begin{aligned}
d_{s+1}&\lesssim d_{s}\binom{n_{s+1}}{k_{s+1}-1}(Q+d_{s})^{k_{s+1}-1}+\binom{n_{s+1}}{k_{s+1}}(Q+d_{s})^{k_{s+1}}
\\
&=(Q+d_{s})^{k_{s+1}-1}[(d_{s}\binom{n_{s+1}}{k_{s+1}-1}+\binom{n_{s+1}}{k_{s+1}}(Q+d_{s})]
\end{aligned}\end{equation}
This expression provides a practical estimation for the DCR suppression at the iterative level $s+1$.

\qquad Similarly, the DE can be lower-bounded by:
\begin{equation}\label{EDE}\begin{aligned}
\eta_{s+1}\geq& p^{n_{s+1}}[P'_{s+1}\sum_{j\geq k_{s+1}-1}\binom{n_{s+1}}{j}P_{s+1}^{j}(1-P_{s+1})^{n_{s+1}-j}
\\
&\qquad\qquad\qquad\qquad+(1-P'_{s+1})\sum_{j\geq k_{s+1}}\binom{n_{s+1}}{j}P_{s+1}^{j}(1-P_{s+1})^{n_{s+1}-j}]
\\
=&p^{n_{s+1}}[P'_{s+1}\binom{n_{s+1}}{k_{s+1}-1}P_{s+1}^{k_{s+1}-1}(1-P_{s+1})^{n_{s+1}-k_{s+1}+1}
\\
&\qquad\qquad\qquad\qquad\qquad\qquad\ +\sum_{j\geq k_{s+1}}\binom{n_{s+1}}{j}P_{s+1}^{j}(1-P_{s+1})^{n_{s+1}-j}]
\\
\approx &p^{n_{s+1}}[\eta_{s}\binom{n_{s+1}}{k_{s+1}-1}(P\eta_{s})^{k_{s+1}-1}(1-P\eta_{s})^{n_{s+1}-k_{s+1}+1}
\\
&\qquad\qquad\qquad\qquad\qquad\qquad\ +\sum_{j\geq k_{s+1}}\binom{n_{s+1}}{j}(P\eta_{s})^{j}(1-P\eta_{s})^{n_{s+1}-j}]
\end{aligned}\end{equation}
This bound is independent of $d_{s}$. Hence, the condition for DE enhancement at the subsequent level (namely $\eta_{s+1} > \eta_s$) is determined by the inequality:
\begin{equation}\label{IDE}\begin{aligned}
p^{n}(Px)^{k-1}[x\binom{n}{k-1}(1-Px)^{n-k+1}+\sum_{j\geq k}\binom{n}{j}(Px)^{j-k+1}(1-Px)^{n-j}]-x>0
\end{aligned}\end{equation}
If $n_s$ and $k_s$ are constant across levels and the system is stable, the DE of the $\text{ESPD}_{s}$ will (approximately) converge to a stable fixed point corresponding to a root of the left-hand side of Eq.~\eqref{IDE}.

\subsection{Monte Carlo Steady-State Analysis and Numerical Simulations}
\label{sec:Experiments}

While obtaining an exact analytical solution for the non-linear dynamic system governing the ESPD in Eq.~\eqref{DynamicSystem} is mathematically complex, numerical methods provide a robust approach for stability analysis and practical design. In this section, we present the Monte Carlo analysis to characterize the system's evolutionary behavior, including the location of stable fixed points and their basins of attraction, and provide numerical simulations under practically motivated device parameters. The results indicate that substantial performance gains can be achieved even with moderate initial hardware specifications. For instance, a three-level ESPD can elevate a conventional semiconductor SPD to effective performance metrics (DE $>93\%$ and DCR $<10^{-9}$), matching state-of-the-art SNSPDs, but without requiring cryogenic infrastructure.

\textbf{Simulation Parameters:} The parameter settings are physically motivated by commercial specifications or experimental reports. The controlled-gate parameters are set to $p=0.98$ and $P=0.97$, values recently demonstrated in optical platforms and feedforward C-NOT gate implementations on photonic platforms~\cite{KH2021Room,HE2023Efficient,SD2025Experimental,AD2025A,P2025A}. For the auxiliary path, the error rate is set to $Q=0.2\%$, justified by the extremely low SPAM errors achievable across specific DOF and the inherent design freedom in selecting the appropriate DOF for the ESPD implementation~\cite{HA2014High,P2025A}. Detailed discussions on parameters and robustness under degraded settings are provided in Section~\ref{sec:Device}.

\begin{figure}[htbp]
    \centering
    \begin{subfigure}[htbp]{0.48\columnwidth}
    \includegraphics[width=\linewidth]{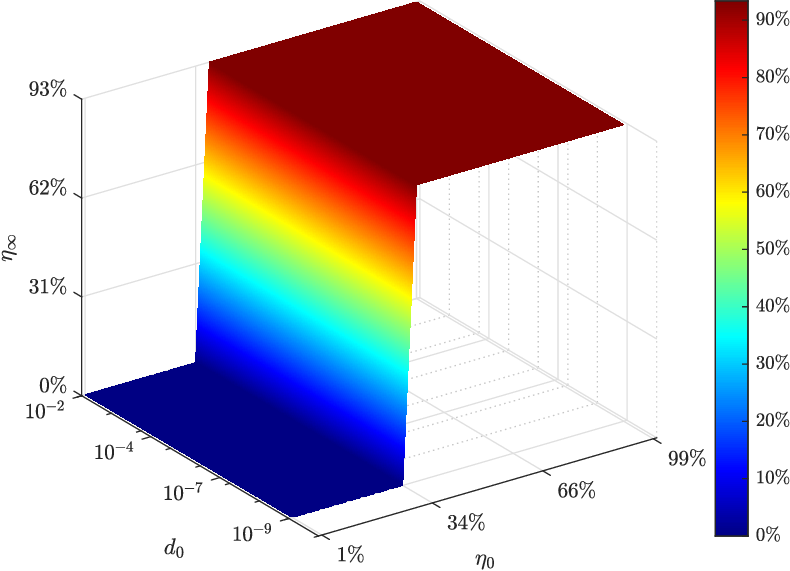}
    \caption{Steady-state DE $\eta_{\infty}$ vs.\ initial DE and DCR $(\eta_{0},d_{0})$.}
    \label{MTKLSS84DE}
    \end{subfigure}
    \hfill
    \begin{subfigure}[htbp]{0.48\columnwidth}
    \includegraphics[width=\linewidth]{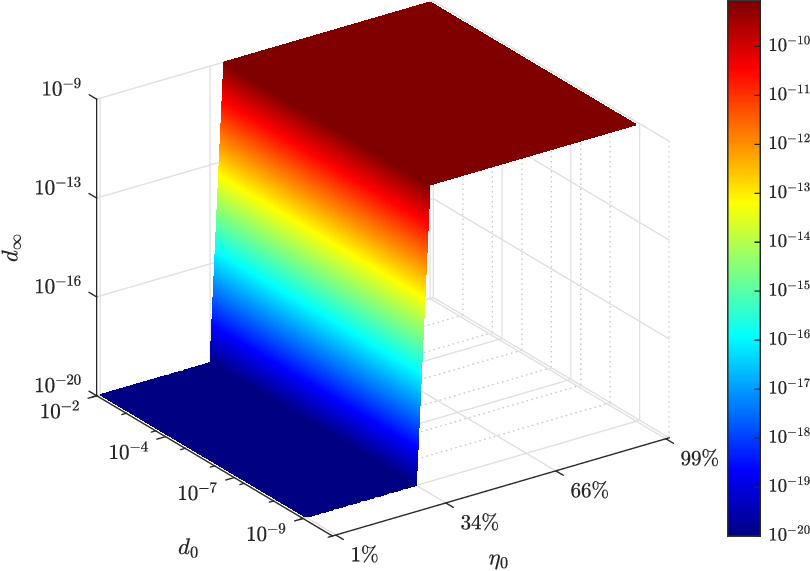}
    \caption{Steady-state DCR $d_{\infty}$ vs.\ initial DE and DCR $(\eta_{0},d_{0})$.}
    \label{MTKLSS84DCR}
    \end{subfigure}
   \caption{\ Monte Carlo steady-state analysis of DE and DCR of the ESPD framework for $(n_{s},k_{s})=(8,4)$ at each level $s$.}
    \label{MTCLSS84}
\end{figure}

\begin{figure}[htbp]
    \centering
    \begin{subfigure}[htbp]{0.48\columnwidth}
    \includegraphics[width=\linewidth]{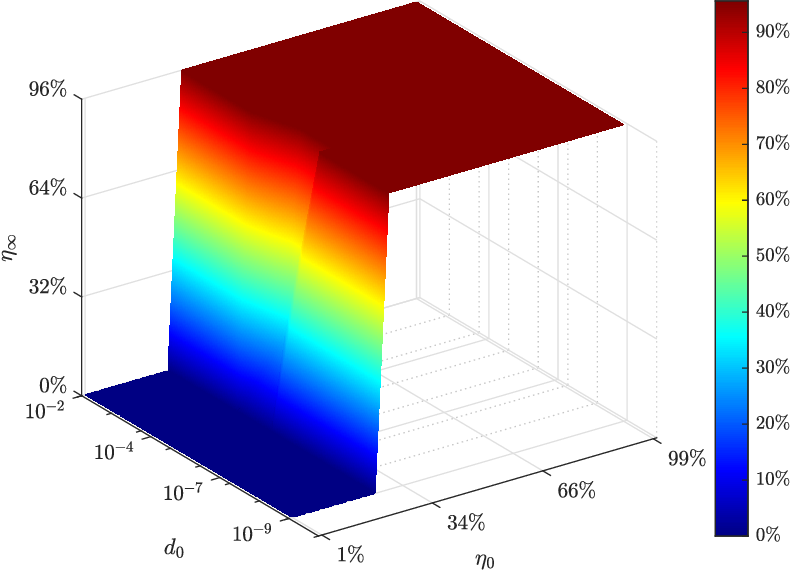}
    \caption{Steady-state DE $\eta_{\infty}$ vs.\ initial DE and DCR $(\eta_{0},d_{0})$.}
    \label{MTKLSS63DE}
    \end{subfigure}
    \hfill
    \begin{subfigure}[htbp]{0.48\columnwidth}
    \includegraphics[width=\linewidth]{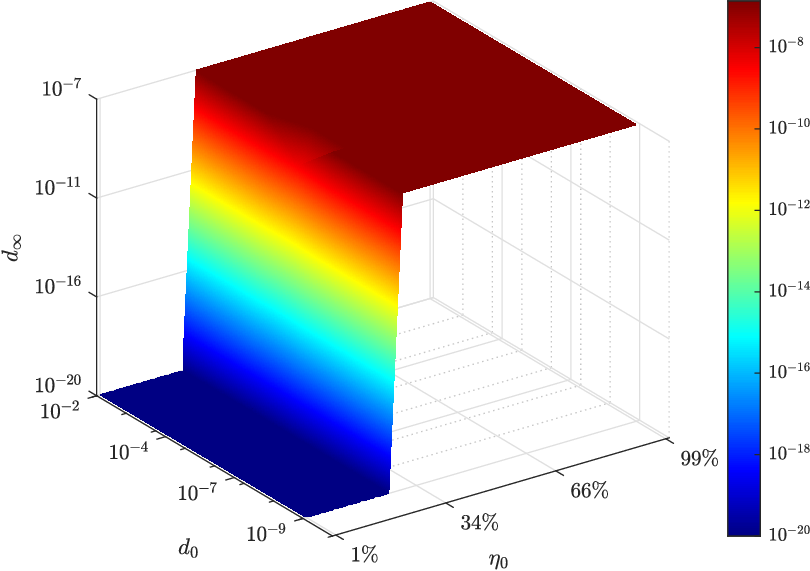}
    \caption{Steady-state DCR $d_{\infty}$ vs.\ initial DE and DCR $(\eta_{0},d_{0})$.}
    \label{MTKLSS63DCR}
    \end{subfigure}
   \caption{\ Monte Carlo steady-state analysis of DE and DCR of the ESPD framework for $(n_{s},k_{s})=(6,3)$ at each level $s$.}
    \label{MTCLSS63}
\end{figure}

\begin{figure}[htbp]
    \centering
    \begin{subfigure}[htbp]{0.48\columnwidth}
    \includegraphics[width=\linewidth]{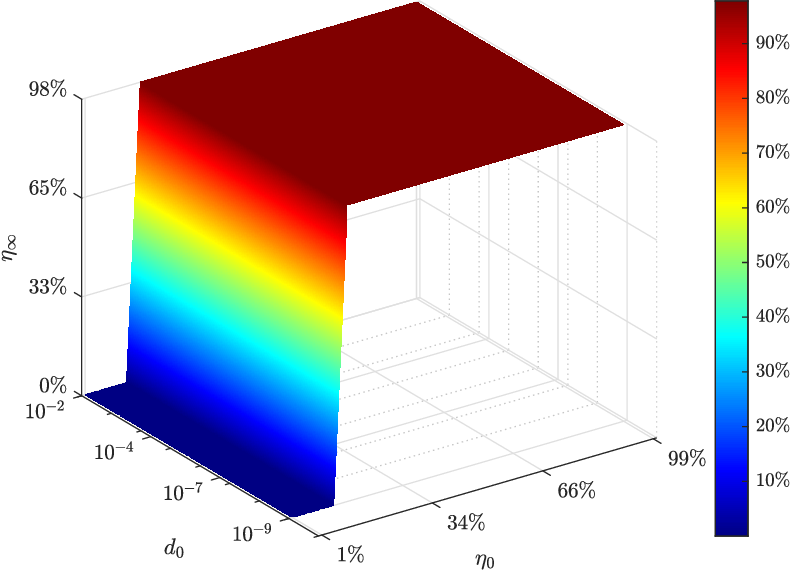}
    \caption{Steady-state DE $\eta_{\infty}$ vs.\ initial DE and DCR $(\eta_{0},d_{0})$.}
    \label{MTKLSS42DE}
    \end{subfigure}
    \hfill
    \begin{subfigure}[htbp]{0.48\columnwidth}
    \includegraphics[width=\linewidth]{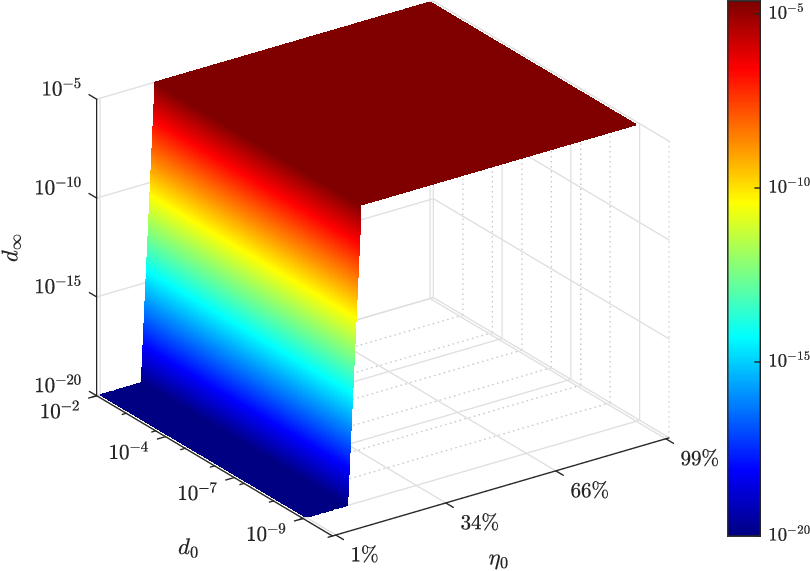}
    \caption{Steady-state DCR $d_{\infty}$ vs.\ initial DE and DCR $(\eta_{0},d_{0})$.}
    \label{MTKLSS42DCR}
    \end{subfigure}
   \caption{\ Monte Carlo steady-state analysis of DE and DCR of the ESPD framework for $(n_{s},k_{s})=(4,2)$ at each level $s$.}
    \label{MTCLSS42}
\end{figure}

\textbf{Monte Carlo Outcomes for Steady-State Analysis:} We conduct Monte Carlo analysis to identify the basins of attraction and the corresponding steady states $(\eta_{\infty}, d_{\infty})$ under various $(n,k)$ configurations. The initial DE and DCR range from 0.01 to 0.99 and $10^{-2}$ to $10^{-9}$, respectively, covering the performance spectrum of most contemporary detectors. Fig.\ref{MTCLSS84}, \ref{MTCLSS63}, and \ref{MTCLSS42} demonstrate that, provided the initial SPD state is not excessively poor, the system converges to a high-performance steady state with appropriate $(n,k)$ values. Specifically, for $(n,k)=(4,2)$, the framework reaches a steady-state DE of $97.80\%$ and DCR of $2.4\times 10^{-5}$ if the initial DE is at least 16\%, and for $(n,k)=(6,3)$, the system converges to a DE of approximately 95.60\% and a DCR of 1.4$\times10^{-7}$ if the initial DE is at least 30\%, while for $(n,k)=(8,4)$, the system converges to a DE of approximately 93.35\% and a DCR of 8.5$\times10^{-10}$ if the initial DE is at least 37\%.

\begin{figure}[htbp]
    \centering
    \begin{subfigure}[htbp]{0.48\columnwidth}
    \includegraphics[width=\linewidth]{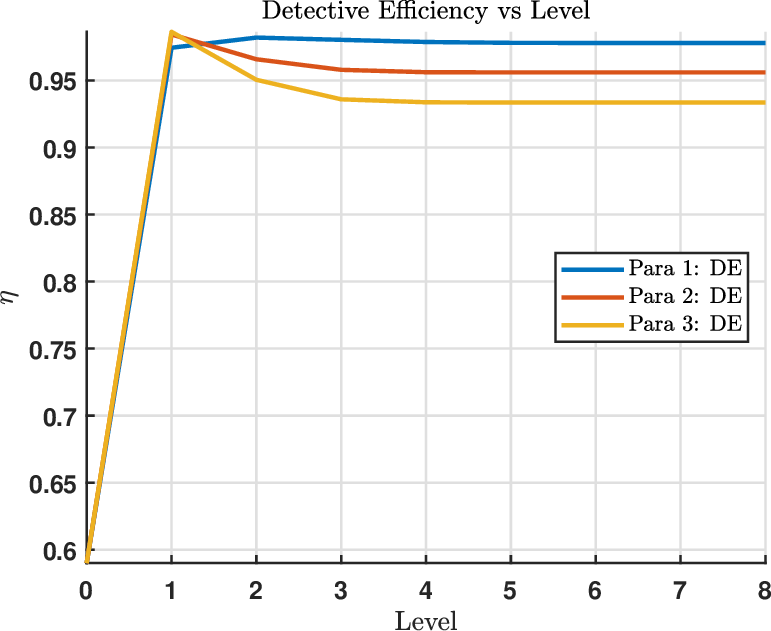}
    \caption{DE $\eta_{s}$ vs.\ level $s$ for varying $(n_{s},k_{s})$.}
    \label{FDE59P97DE}
    \end{subfigure}
    \hfill
    \begin{subfigure}[htbp]{0.48\columnwidth}
    \includegraphics[width=\linewidth]{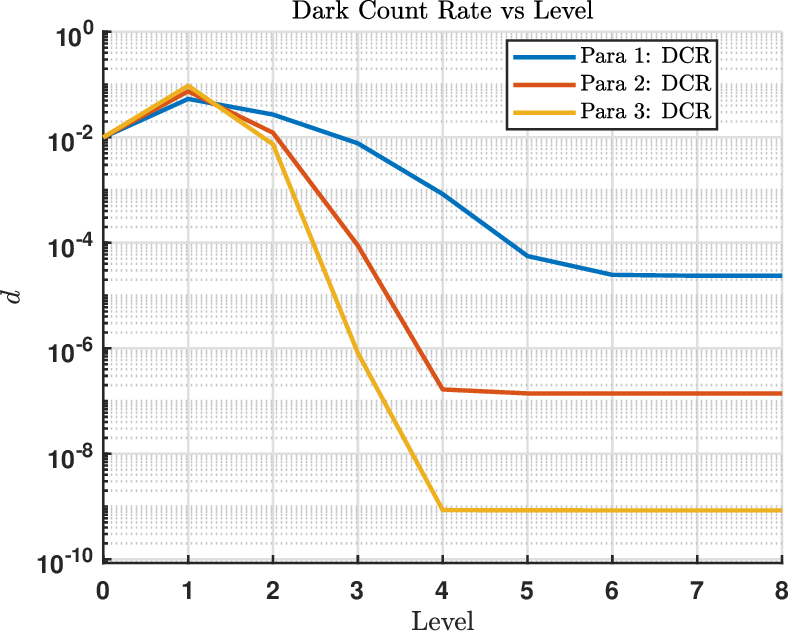}
    \caption{DCR $d_{s}$ vs.\ level $s$ for varying $(n_{s},k_{s})$.}
    \label{FDE59P97DCR}
    \end{subfigure}
   \caption{\ Performance of the ESPD initialized with SPDs from \cite{AW2004Efficient} with $\eta_{0}=59.0\%$ and $d_{0}=1.0\times 10^{-2}$.}
    \label{FDE59P97}
\end{figure}

\textbf{Performance Evolution on Legacy SPD:} We specifically examine the evolutionary trajectory of the ESPD initialized with a legacy SPD from~\cite{AW2004Efficient} (DE $\approx 59\%$, DCR $\approx 10^{-2}$). Table~\ref{TDE59P97} details the step-by-step evolutions, with visual representations in Fig.~\ref{FDE59P97}. These results illustrate that the ESPD scheme significantly transcends the intrinsic limitations of conventional SPDs. Notably, within only three to four levels of iterative enhancement, the effective metrics (DE $> 93\%$, DCR $< 10^{-9}$) become comparable to advanced superconducting SPDs, consistent with the stability predicted by our steady-state analysis.

\begin{table}[htbp]
\renewcommand\arraystretch{1}
\caption{Performance of an ESPD initialized with an SPD from \cite{AW2004Efficient}, with $\eta_{0}=59\%,\ d_{0}=10^{-2}$.}
\centering
\label{TDE59P97}
\footnotesize
\begin{tabular}{@{\hspace{0.1mm}}c@{\hspace{1.5mm}}c@{\hspace{1mm}}c@{\hspace{1.5mm}}c@{\hspace{1mm}}c@{\hspace{1.5mm}}c@{\hspace{1mm}}c@{\hspace{0.1mm}}}
\hline
 & \multicolumn{2}{c}{Para 1: $(n,k)=(4,2)$} & \multicolumn{2}{c}{Para 2: $(n,k)=(6,3)$} & \multicolumn{2}{c}{Para 3: $(n,k)=(8,4)$}\\
\hline
Level & $(n,k)$ & (DE\ ($\eta$),\ DCR\ ($d$)) & $(n,k)$ & (DE\ ($\eta$),\ DCR\ ($d$)) & $(n,k)$ & (DE\ ($\eta$),\ DCR\ ($d$))\\
\hline
0 & - & $(59.0\%,\ 1.0\times 10^{-2})$ & - & $(59.0\%,\ 1.0\times 10^{-2})$ & - & $(59.0\%,\ 1.0\times 10^{-2})$\\
1 & (4,2) & $(97.4\%,\ 5.3\times 10^{-2})$ & (6,3) & $(98.4\%,\ 7.5\times 10^{-2})$ & (8,4) & $(98.6\%,\ 9.5\times 10^{-2})$\\
2 & (4,2) & $(98.2\%,\ 2.7\times 10^{-2})$ & (6,3) & $(96.6\%,\ 1.2\times 10^{-2})$ & (8,4) & $(95.1\%,\ 7.4\times 10^{-3})$\\
3 & (4,2) & $(98.0\%,\ 7.7\times 10^{-3})$ & (6,3) & $(95.8\%,\ 8.9\times 10^{-5})$ & (8,4) & $(93.6\%,\ 8.1\times 10^{-7})$\\
4 & (4,2) & $(97.9\%,\ 8.4\times 10^{-4})$ & (6,3) & $(95.6\%,\ 1.7\times 10^{-7})$ & (8,4) & $(93.4\%,\ 8.6\times 10^{-10})$\\
5 & (4,2) & $(97.8\%,\ 5.6\times 10^{-5})$ & (6,3) & $(95.6\%,\ 1.4\times 10^{-7})$ & (8,4) & $(93.4\%,\ 8.5\times 10^{-10})$\\
6 & (4,2) & $(97.8\%,\ 2.5\times 10^{-5})$ & (6,3) & $(95.6\%,\ 1.4\times 10^{-7})$ & (8,4) & $(93.4\%,\ 8.5\times 10^{-10})$\\
7 & (4,2) & $(97.8\%\ ,2.4\times 10^{-5})$ & (6,3) & $(95.6\%,\ 1.4\times 10^{-7})$ & (8,4) & $(93.4\%,\ 8.5\times 10^{-10})$\\
8 & (4,2) & $(97.8\%,\ 2.4\times 10^{-5})$ & (6,3) & $(95.6\%,\ 1.4\times 10^{-7})$ & (8,4) & $(93.4\%,\ 8.5\times 10^{-10})$\\
\hline
\end{tabular}
\end{table}

\subsection{Implications for QKD}
\label{sec:Application}

As a representative application, we analyze the implications of the ESPD framework applied to QKD. 

In a QKD protocol characterized by a secure error threshold $e_{th}$ and a baseline error rate $e_{C}$ (excluding detector-induced errors), the minimal tolerable channel transmission rate $\gamma$, which defines the operational limit of a link, is conventionally approximated by\cite{S2025Elimiating}:
\begin{equation}\begin{aligned}
\gamma=\frac{(1-2e_{th})d}{\eta[e_{th}-e_{C}+d(1-2e)]}\approx \frac{1-2e_{th}}{e_{th}-e_{C}}\frac{d}{\eta}
\end{aligned}\end{equation}

Since the ESPD paradigm can substantially suppress effective DCR $d$ while simultaneously enhancing the DE $\eta$, it enables a dramatic reduction in $\gamma$. This suppression of the minimal transmission threshold directly translates into an extended communication distance and increased robustness against channel loss\footnote{The channel transmission rate $t$ is conventionally calculated by $t=10^{-\frac{\alpha l}{10}}$, where $\alpha$ is a fixed parameter mainly determined by channel characters and $l$ is the communication distance.}. Furthermore, the enhancement of $\eta$ not only lowers the threshold $\gamma$ but also scales the achievable secret key rate at any fixed channel transmission, significantly improving the system throughput.

Compared to existing alternative strategies, such as the empty-signal-detection paradigm~\cite{S2025Elimiating}, the ESPD framework offers a distinct functional advantage. While the previous method also reduces the channel transmission threshold, it introduces additional sifting consumption that inherently degrades the efficiency. In contrast, the ESPD framework achieves performance gains by directly improving DE and reducing DCR, without the need for auxiliary sifting. Consequently, the ESPD paradigm provides a more efficient approach for improving QKD performance under practical constraints.

\section{Discussion}

In this section, we discuss the technological feasibility and practical robustness of the ESPD framework, as well as its limitations and directions for future work.

\subsection{Effectiveness when the Initial SPD is Outside the Convergent Range}

As evidenced by the Monte Carlo analysis, the steady-state performance of the ESPD is largely independent of the specific characteristics of the initial SPD, provided that it lies within a convergence basin. However, although the convergence interval is relaxed, some traditional SPDs might still fall outside this range. For example, the InGaAs/InP SPD designed in 2017 with a relatively lower DE ($\approx 27.5\%$) and lower DCR ($\approx 10^{-6}$)\cite{JL20171.25} is outside the range in which the ESPD dynamics converge to a high-performance fixed point when a uniform configuration of $(n_{s},k_{s})=(6,3)$ or $(n_{s},k_{s})=(8,4)$ is shared by each level. 

Nevertheless, such SPDs can still be significantly enhanced by the ESPD framework through an appropriate selection of the first-level configuration parameters $(n_{1},k_{1})$. This strategic parameter choice allows states of the dynamical system described by Eq.~\eqref{DynamicSystem} to be driven into the basin of attraction at the first step. For instance, selecting $(n_{1},k_{1})=(6,1)$ and $(n_{1},k_{1})=(8,1)$ results in $(\eta_{1},d_{1})=(85.5\%,3.4\times 10^{-3})$ and $(\eta_{1},d_{1})=(90.0\%,4.5\times 10^{-3})$, respectively, starting from the aforementioned SPD from \cite{JL20171.25}, which successfully falls within the convergence basin for subsequent levels. Indeed, while a trade-off exists between DE and DCR, DCR of the ESPD can be drastically reduced by increasing $k$, and DE can be significantly enhanced by selecting a small $k$. Therefore, by employing one or two levels of the ESPD paradigm with suitable configuration parameters as a buffer, the detector's performance can be adjusted into a desirable operational window, as long as the initial SPD is not prohibitively low. 

\begin{figure}[htbp]
    \centering
    \begin{subfigure}[htbp]{0.48\columnwidth}
    \includegraphics[width=\linewidth]{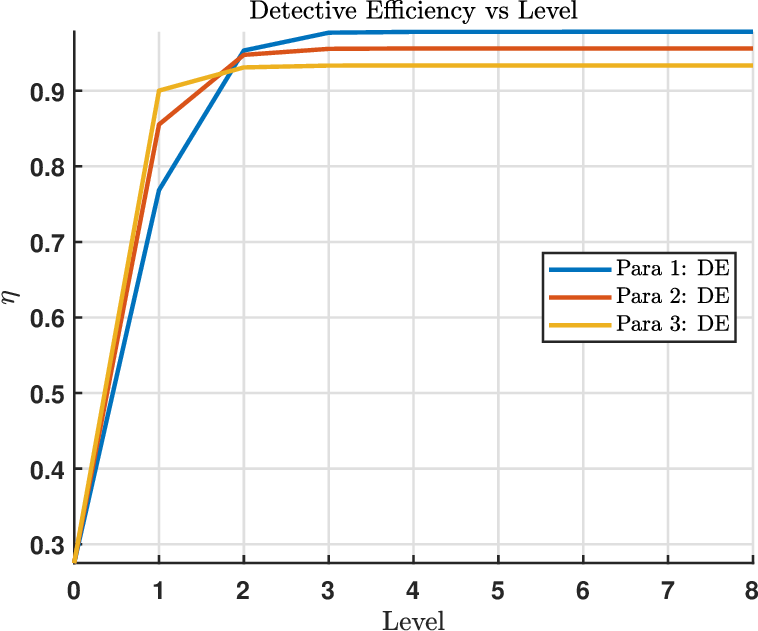}
    \caption{DE $\eta_{s}$ vs.\ level $s$ for varying $(n_{s},k_{s})$.}
    \label{FDE27P97DE}
    \end{subfigure}
    \hfill
    \begin{subfigure}[htbp]{0.48\columnwidth}
    \includegraphics[width=\linewidth]{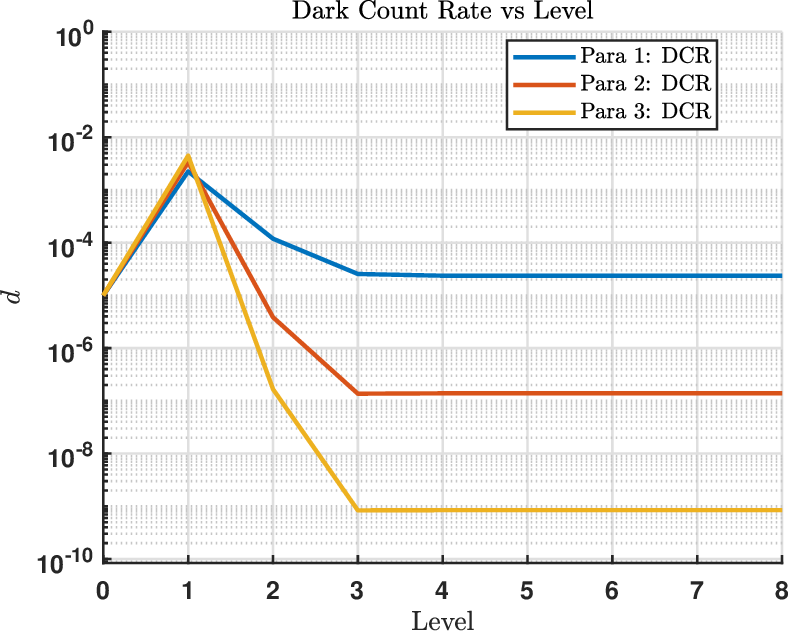}
    \caption{DCR $d_{s}$ vs.\ level $s$ for varying $(n_{s},k_{s})$.}
    \label{FDE27P97DCR}
    \end{subfigure}
   \caption{\ Performance of ESPD initialized with an SPD from \cite{JL20171.25} with initial parameters: $\eta_{0}=27.5\%$ and $d_{0}=1.0\times 10^{-6}$, respectively.}
    \label{FDE27P97}
\end{figure}

Numerical simulations for an initial SPD with $(\eta_{0},d_{0})=(27.5\%,1.0\times 10^{-6})$ are provided in Table \ref{TDE27P97} and visualized in Fig.\ref{FDE27P97}, suggesting that SPDs with either low DE and low DCR~\cite{JL20171.25}, or moderate DE and high DCR~\cite{AW2004Efficient}, can serve as viable initializations.

\begin{table}[htbp]
\renewcommand\arraystretch{1}
\caption{Performance of ESPD initialized with an SPD from \cite{JL20171.25}, with initial parameters $\eta_{0}=27.5\%,\ d_{0}=10^{-6}$.}
\centering
\label{TDE27P97}
\footnotesize
\begin{tabular}{@{\hspace{0.1mm}}c@{\hspace{1.5mm}}c@{\hspace{1mm}}c@{\hspace{1.5mm}}c@{\hspace{1mm}}c@{\hspace{1.5mm}}c@{\hspace{1mm}}c@{\hspace{0.1mm}}}
\hline
 & \multicolumn{2}{c}{Para 1} & \multicolumn{2}{c}{Para 2} & \multicolumn{2}{c}{Para 3}\\
\hline
Level & $(n,k)$ & (DE\ ($\eta$),\ DCR\ ($d$)) & $(n,k)$ & (DE\ ($\eta$),\ DCR\ ($d$)) & $(n,k)$ & (DE\ ($\eta$),\ DCR\ ($d$))\\
\hline
0 & - & $(27.5\%,\ 1.0\times 10^{-6})$ & - & $(27.5\%,\ 1.0\times 10^{-6})$ & - & $(27.5\%,\ 1.0\times 10^{-6})$\\
1 & (4,1) & $(76.9\%,\ 2.2\times 10^{-3})$ & (6,1) & $(85.5\%,\ 3.4\times 10^{-3})$ & (8,1) & $(90.0\%,\ 4.5\times 10^{-3})$\\
2 & (4,2) & $(95.3\%,\ 1.2\times 10^{-4})$ & (6,2) & $(94.8\%,\ 3.8\times 10^{-4})$ & (8,4) & $(93.1\%,\ 1.7\times 10^{-7})$\\
3 & (4,2) & $(97.7\%,\ 2.6\times 10^{-5})$ & (6,3) & $(95.5\%,\ 1.4\times 10^{-7})$ & (8,4) & $(93.3\%,\ 8.4\times 10^{-10})$\\
4 & (4,2) & $(97.8\%,\ 2.4\times 10^{-5})$ & (6,3) & $(95.6\%,\ 1.4\times 10^{-7})$ & (8,4) & $(93.4\%,\ 8.4\times 10^{-10})$\\
5 & (4,2) & $(97.8\%,\ 2.4\times 10^{-5})$ & (6,3) & $(95.6\%,\ 1.4\times 10^{-7})$ & (8,4) & $(93.4\%,\ 8.5\times 10^{-10})$\\
6 & (4,2) & $(97.8\%,\ 2.4\times 10^{-5})$ & (6,3) & $(95.6\%,\ 1.4\times 10^{-7})$ & (8,4) & $(93.4\%,\ 8.5\times 10^{-10})$\\
7 & (4,2) & $(97.8\%,\ 2.4\times 10^{-5})$ & (6,3) & $(95.6\%,\ 1.4\times 10^{-7})$ & (8,4) & $(93.4\%,\ 8.5\times 10^{-10})$\\
8 & (4,2) & $(97.8\%,\ 2.4\times 10^{-5})$ & (6,3) & $(95.6\%,\ 1.4\times 10^{-7})$ & (8,4) & $(93.4\%,\ 8.5\times 10^{-10})$\\
\hline
\end{tabular}
\end{table}

\subsection{Technological Feasibility and Robustness under Degraded Parameters}
\label{sec:Device}

The primary technological advantage of the ESPD paradigm lies in its exclusive reliance on accessible, room-temperature components, effectively circumventing the stringent cryogenic requirements inherent to superconducting SPDs. The technological feasibility of components is discussed as follows.

\subsubsection{State Preparation and Measurement}

The architectural flexibility of the ESPD framework allows for various DOFs to be employed, enabling the selection of an easily manipulated DOF based on specific platforms. In practical implementations, this facilitates high-fidelity SPAM, ensuring that the error parameter $Q$ can be maintained at a low level. Specifically, $Q$ can be suppressed to the $10^{-4}$ level using standard optical setups and commercial platforms \cite{HA2014High,P2025A}, guaranteeing $Q\ll 1$.

\subsubsection{Controlled Gates}
\label{sec:ControlledGate}

The primary technological prerequisites of the ESPD scheme are the controlled quantum gates, typically C-NOT operations, which define the critical parameters $p$ and $P$.

High transmission efficiencies, for example, with attenuation as low as $0.1$ dB (corresponding to a propagation probability $p \approx 0.98$) \cite{P2025A}, are achievable with commercial photonic integrated circuits \cite{MD2005Propagation, CL2022Ultra, FZ2022Transverse, FZ2025Chip}, suggesting that $p$ can approach unity. It is important to reiterate that $p$ represents the intrinsic transmission rate of a single, non-cascaded controlled module and, by definition, remains invariant under an increasing number of cascaded modules or ESPD levels. The cumulative system loss is fully accounted for in the derivation of the total system efficiency, $\eta_{s+1}$ (Eq.\eqref{DE}).

\begin{table}[htbp]
\renewcommand\arraystretch{1}
\caption{Performance of an ESPD initialized with an SPD from \cite{AW2004Efficient}, with $p$ varied from 0.80 to 0.96. Simulations are performed for $(n,k)=(5,2)$ at each level $L$, and all other settings remain consistent with those described in Section \ref{sec:Experiments}.}
\centering
\scriptsize
\label{TDE59Pvar}
\begin{tabular}{@{\hspace{0.1mm}}c@{\hspace{0.5mm}}c@{\hspace{0.5mm}}c@{\hspace{0.5mm}}c @{\hspace{0.5mm}}c@{\hspace{0.5mm}}c@{\hspace{0.1mm}}}
\hline
 & $p$=0.80 & $p$=0.84 & $p$=0.88 & $p$=0.92 & $p$=0.96\\
\hline
$L$ & (DE\ ($\eta$),\ DCR\ ($d$)) & (DE\ ($\eta$),\ DCR\ ($d$)) & (DE\ ($\eta$),\ DCR\ ($d$)) & (DE\ ($\eta$),\ DCR\ ($d$)) & (DE\ ($\eta$),\ DCR\ ($d$))\\
0 & $(59.0\%,1.0\times 10^{-2})$ & $(59.0\%,1.0\times 10^{-2})$ & $(59.0\%,1.0\times 10^{-2})$ & $(59.0\%,1.0\times 10^{-2})$ & $(59.0\%,1.0\times 10^{-2})$ \\
1 & $(60.8\%,1.8\times 10^{-3})$ & $(66.7\%,1.8\times 10^{-3})$ & $(72.9\%,1.8\times 10^{-3})$ & $(79.7\%,1.8\times 10^{-3})$  & $(87.0\%,1.8\times 10^{-3})$\\
2 & $(61.0\%,1.1\times 10^{-4})$ & $(70.6\%,1.2\times 10^{-4})$ & $(79.7\%,1.3\times 10^{-4})$ & $(87.8\%,1.4\times 10^{-4})$  & $(94.6\%,1.5\times 10^{-4})$\\
3 & $(61.0\%,1.8\times 10^{-5})$ & $(72.6\%,2.4\times 10^{-5})$ & $(82.1\%,3.1\times 10^{-5})$ & $(89.5\%,3.7\times 10^{-5})$  & $(95.4\%,4.3\times 10^{-5})$\\
4 & $(60.9\%,1.5\times 10^{-5})$ & $(73.5\%,2.2\times 10^{-5})$ & $(82.8\%,2.8\times 10^{-5})$ & $(89.9\%,3.4\times 10^{-5})$  & $(95.4\%,3.8\times 10^{-5})$\\
5 & $(60.8\%,1.5\times 10^{-5})$ & $(74.0\%,2.2\times 10^{-5})$ & $(83.1\%,2.9\times 10^{-5})$ & $(89.9\%,3.4\times 10^{-5})$  & $(95.4\%,3.8\times 10^{-5})$\\
6 & $(60.8\%,1.5\times 10^{-5})$ & $(74.2\%,2.3\times 10^{-5})$ & $(83.2\%,2.9\times 10^{-5})$ & $(89.9\%,3.4\times 10^{-5})$  & $(95.4\%,3.8\times 10^{-5})$\\
7 & $(60.8\%,1.5\times 10^{-5})$ & $(74.3\%,2.3\times 10^{-5})$ & $(83.2\%,2.9\times 10^{-5})$ & $(89.9\%,3.4\times 10^{-5})$  & $(95.4\%,3.8\times 10^{-5})$\\
8 & $(60.8\%,1.5\times 10^{-5})$ & $(74.3\%,2.3\times 10^{-5})$ & $(83.2\%,2.9\times 10^{-5})$ & $(89.9\%,3.4\times 10^{-5})$  & $(95.4\%,3.8\times 10^{-5})$\\
\hline
\end{tabular}
\end{table}

\begin{figure}[htbp]
    \centering
    \begin{subfigure}[htbp]{0.48\columnwidth}
    \includegraphics[width=\linewidth]{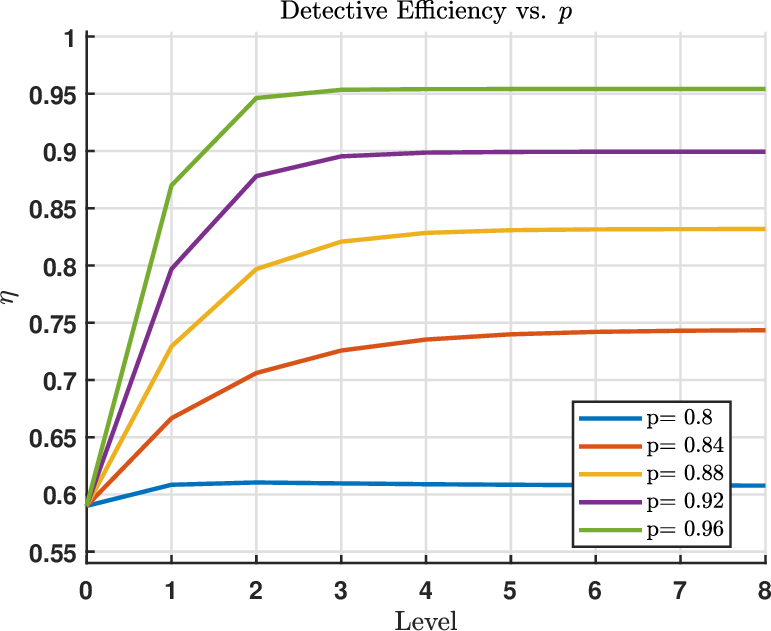}
    \caption{DE $\eta_{s}$ vs.\ $p$ for $(n_{s},k_{s})=(5,2)$.}
    \end{subfigure}
    \begin{subfigure}[htbp]{0.48\columnwidth}
    \includegraphics[width=\linewidth]{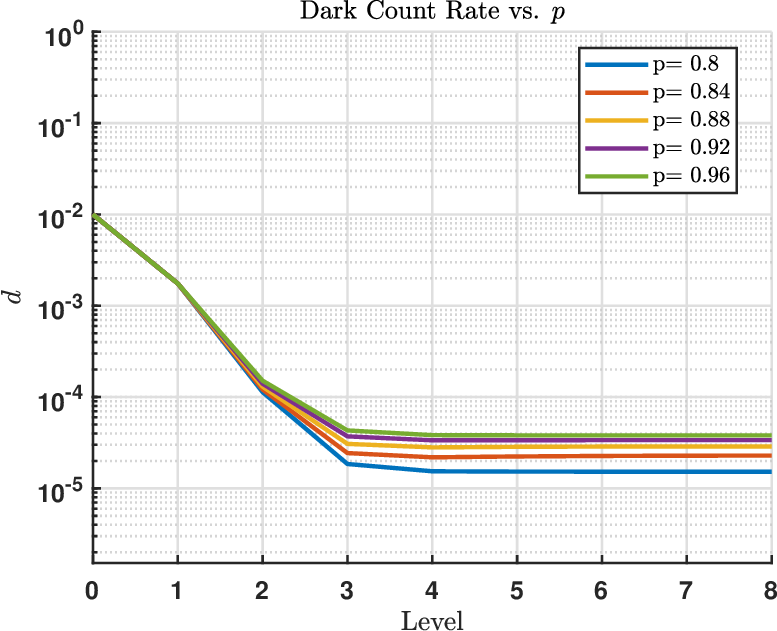}
    \caption{DCR $d_{s}$ vs.\ level $p$ for $(n_{s},k_{s})=(5,2)$.}
    \end{subfigure}
    \caption{\ Performance for parameters $\eta_{0}=59.0\%$, $d_{0}=1.0\times 10^{-2}$ with degraded $p$. All other settings remain consistent with those described in Section \ref{sec:Experiments}.}
    \label{FDE59Pvar}
\end{figure}

However, to accommodate practical implementations, one may consider using post-selection controlled gates instead of the feedforward configurations. While the ESPD paradigm remains valid for such probabilistic operations (with slight modifications in analysis), the overall system efficiency is necessarily degraded. Specifically, if the post-selection efficiency of the controlled gate is $\frac{1}{N}$, the effective average transmission rate from one module to the next is reduced to $p^{N}$. Although a full theoretical calculation for the post-selection case is structurally similar to that established in Section \ref{sec:Theory} and is not repeated here, the resulting degradation underscores the importance of characterizing system performance under reduced transmission rate. Table \ref{TDE59Pvar} and Fig.\ref{FDE59Pvar} indicate that the ESPD framework maintains performance enhancement even when $p$ drops to $0.80$ (approximately 1 dB attenuation, corresponding to the loss accumulated across approximately 10 controlled modules with 0.1 dB attenuation on each.). This result shows robustness, indicating the scheme's feasibility even for C-NOT gates with a low post-selection efficiency of around $10\%$, which is well within the reach of current photonic technology \cite{LP2010High, CR2011Integrated, PR2025High, NF2025Quantum, GN2025Quantum, P2025A}.

\begin{figure}[htbp]
    \centering
    \begin{subfigure}[htbp]{0.48\columnwidth}
    \includegraphics[width=\linewidth]{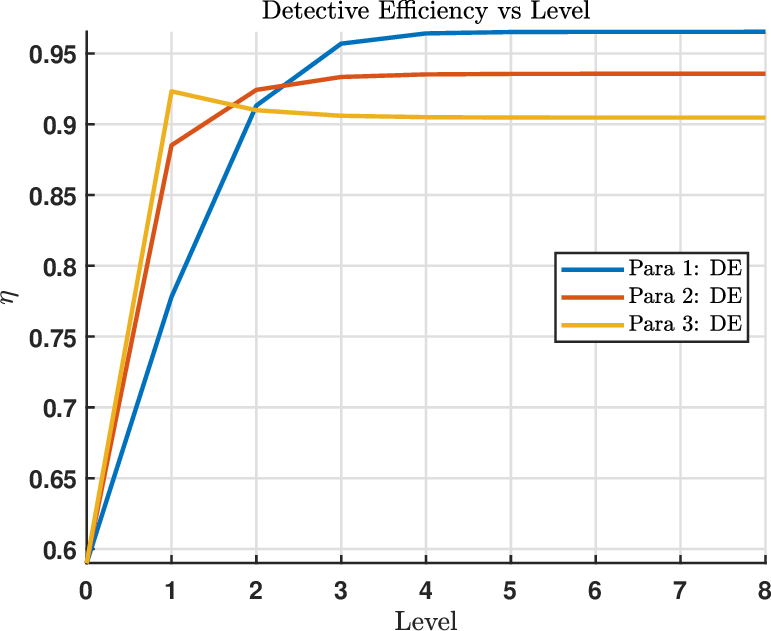}
    \caption{DE $\eta_{s}$ vs.\ level $s$.}
    \end{subfigure}
    \hfill
    \begin{subfigure}[htbp]{0.48\columnwidth}
    \includegraphics[width=\linewidth]{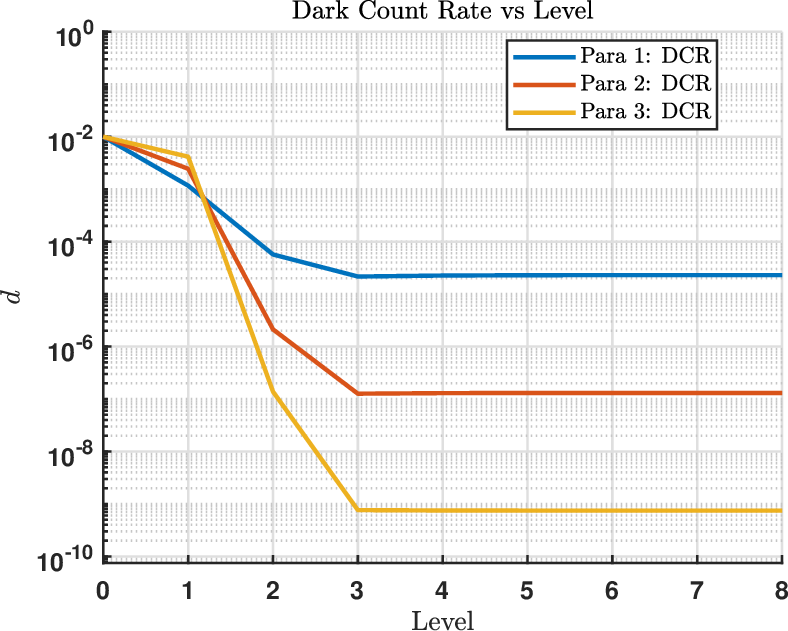}
    \caption{DCR $d_{s}$ vs.\ level $s$.}
    \end{subfigure}
    \caption{\ Performance for ESPD with initialized parameters $\eta_{0}=59.0\%$, $d_{0}=1.0\times 10^{-2}$, with $P=0.80$.}
    \label{FDE59P80}
\end{figure}

\begin{figure}[htbp]
    \centering
    \begin{subfigure}[htbp]{0.48\columnwidth}
    \includegraphics[width=\linewidth]{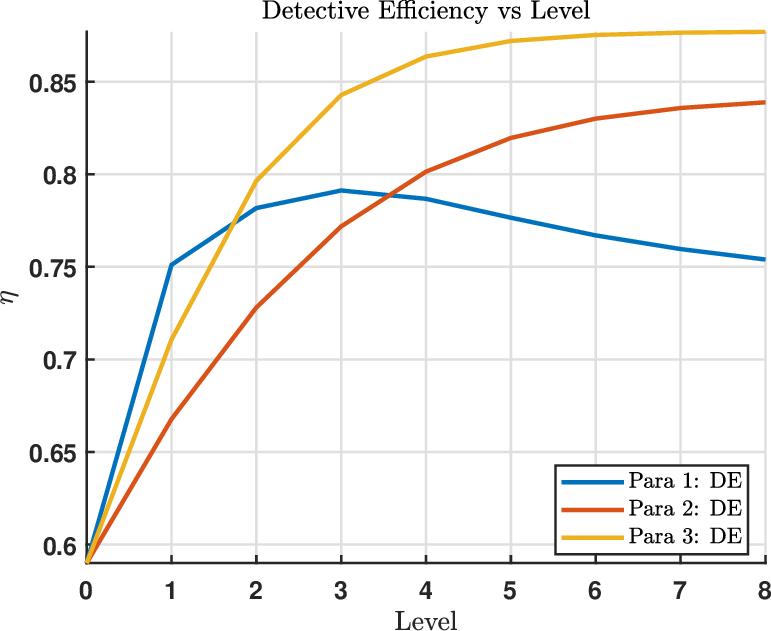}
    \caption{DE $\eta_{s}$ vs.\ level $s$.}
    \end{subfigure}
    \hfill
    \begin{subfigure}[htbp]{0.48\columnwidth}
    \includegraphics[width=\linewidth]{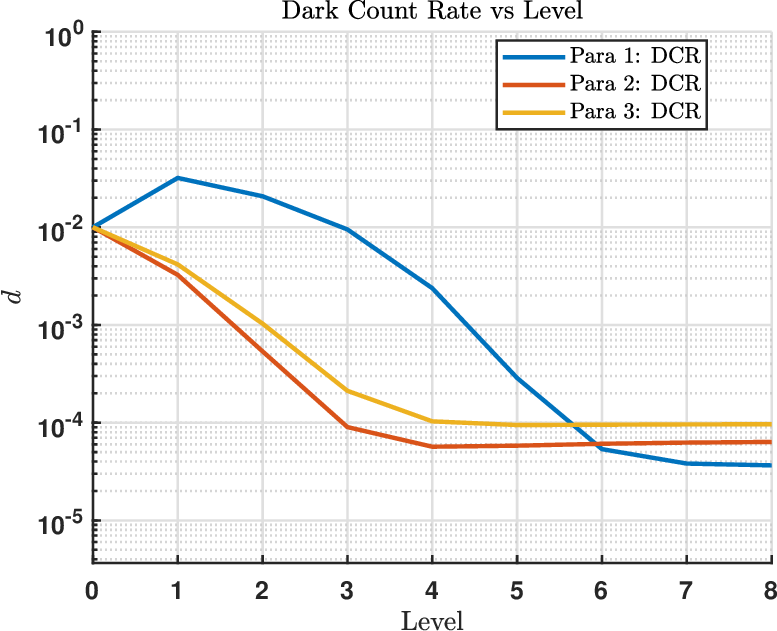}
    \caption{DCR $d_{s}$ vs.\ level $s$.}
    \end{subfigure}
    \caption{\ Performance for ESPD with initialized parameters $\eta_{0}=59.0\%$, $d_{0}=1.0\times 10^{-2}$, with $P=0.40$.}
    \label{FDE59P40}
\end{figure}

We now turn our attention to the parameter $P$, which is intrinsically linked to the fidelity of the C-NOT gate. Notably, $P$ is lower-bounded by the gate's overall fidelity, as $P$ quantifies deviation only on the controlled path, whereas the full C-NOT fidelity should account for the correct operation rate of both the control and controlled paths. Contemporary room-temperature implementations for C-NOT demonstrate high performance, suggesting that high values of $P$ are readily attainable\cite{LP2010High,CR2011Integrated,PR2025High,NF2025Quantum,GN2025Quantum,P2025A}. Specifically, the recent feedforward photonic C-NOT experiment can achieve fidelity over 97\% at non-cryogenic components\cite{KH2021Room,HE2023Efficient,SD2025Experimental,AD2025A}. However, it is also worth further testing the paradigm's resilience under degraded parameters. Here, we also provide simulation results for two degraded scenarios: $P=0.80$ (see Table \ref{TDE59P80}), and $P=0.40$ (see Table \ref{TDE59P40}), whose visual representation is shown in Fig.\ref{FDE59P80} and \ref{FDE59P40}. All other settings remain consistent with those described in Section \ref{sec:Experiments}. The numerical results indicate the robustness of the scheme: Comparison of Table \ref{TDE59P80} with Table \ref{TDE59P97} supports that the ESPD paradigm does not rely on an excessively demanding value for $P$, since similar substantial performance gains are shown for both $P=0.97$ and $P=0.80$, while Table \ref{TDE59P40} implies that the scheme is effective even when the C-NOT fidelity parameter is severely degraded, such that $P=0.40$. This robustness demonstrates that the technological requirements of the ESPD method are fully compatible with and near-term room-temperature quantum fabrication processes.

\begin{table}[htbp]
\renewcommand\arraystretch{1}
\caption{Performance evolution of the ESPD initialized with an SPD from \cite{AW2004Efficient}, and moderate controlled gate fidelity, $P=0.80$.}
\centering
\label{TDE59P80}
\footnotesize
\begin{tabular}{@{\hspace{0.1mm}}c@{\hspace{1.5mm}}c@{\hspace{1mm}}c@{\hspace{1.5mm}}c@{\hspace{1mm}}c@{\hspace{1.5mm}}c@{\hspace{1mm}}c@{\hspace{0.1mm}}}
\hline
 & \multicolumn{2}{c}{Para 1} & \multicolumn{2}{c}{Para 2} & \multicolumn{2}{c}{Para 3}\\
\hline
Level & $(n,k)$ & (DE\ ($\eta$),\ DCR\ ($d$)) & $(n,k)$ & (DE\ ($\eta$),\ DCR\ ($d$)) & $(n,k)$ & (DE\ ($\eta$),\ DCR\ ($d$))\\
\hline
0 & - & $(59.0\%,\ 1.0\times 10^{-2})$ & - & $(59.0\%,\ 1.0\times 10^{-2})$ & - & $(59.0\%,\ 1.0\times 10^{-2})$\\
1 & (4,2) & $(77.8\%,\ 1.2\times 10^{-3})$ & (6,2) & $(88.5\%,\ 2.4\times 10^{-3})$ & (8,2) & $(92.3\%,\ 4.2\times 10^{-3})$\\
2 & (4,2) & $(91.3\%,\ 5.7\times 10^{-5})$ & (6,3) & $(92.4\%,\ 2.1\times 10^{-6})$ & (8,4) & $(91.0\%,\ 1.4\times 10^{-7})$\\
3 & (4,2) & $(95.7\%,\ 2.2\times 10^{-5})$ & (6,3) & $(93.3\%,\ 1.3\times 10^{-7})$ & (8,4) & $(90.6\%,\ 7.6\times 10^{-10})$\\
4 & (4,2) & $(96.4\%,\ 2.3\times 10^{-5})$ & (6,3) & $(93.5\%,\ 1.3\times 10^{-7})$ & (8,4) & $(90.5\%,\ 7.5\times 10^{-10})$\\
5 & (4,2) & $(96.5\%,\ 5.5\times 10^{-5})$ & (6,3) & $(93.6\%,\ 1.3\times 10^{-7})$ & (8,4) & $(90.5\%,\ 7.5\times 10^{-10})$\\
6 & (4,2) & $(96.5\%,\ 2.4\times 10^{-5})$ & (6,3) & $(93.6\%,\ 1.3\times 10^{-7})$ & (8,4) & $(90.5\%,\ 7.5\times 10^{-10})$\\
7 & (4,2) & $(96.5\%,\ 2.3\times 10^{-5})$ & (6,3) & $(93.6\%,\ 1.3\times 10^{-7})$ & (8,4) & $(90.5\%,\ 7.5\times 10^{-10})$\\
8 & (4,2) & $(96.5\%,\ 2.3\times 10^{-5})$ & (6,3) & $(93.6\%,\ 1.3\times 10^{-7})$ & (8,4) & $(90.5\%,\ 7.5\times 10^{-10})$\\
\hline
\end{tabular}
\end{table}

\begin{table}[htbp]
\renewcommand\arraystretch{1}
\caption{Performance evolution of the ESPD initialized with an SPD from \cite{AW2004Efficient}, under a low controlled gate fidelity, $P=0.40$.}
\centering
\label{TDE59P40}
\footnotesize
\begin{tabular}{@{\hspace{0.1mm}}c@{\hspace{1.5mm}}c@{\hspace{1mm}}c@{\hspace{1.5mm}}c@{\hspace{1mm}}c@{\hspace{1.5mm}}c@{\hspace{1mm}}c@{\hspace{0.1mm}}}
\hline
 & \multicolumn{2}{c}{Para 1} & \multicolumn{2}{c}{Para 2} & \multicolumn{2}{c}{Para 3}\\
\hline
Level & $(n,k)$ & (DE\ ($\eta$),\ DCR\ ($d$)) & $(n,k)$ & (DE\ ($\eta$),\ DCR\ ($d$)) & $(n,k)$ & (DE\ ($\eta$),\ DCR\ ($d$))\\
\hline
0 & - & $(59.0\%,\ 1.0\times 10^{-2})$ & - & $(59.0\%,\ 1.0\times 10^{-2})$ & - & $(59.0\%,\ 1.0\times 10^{-2})$\\
1 & (2,1) & $(75.1\%,\ 3.2\times 10^{-2})$ & (7,2) & $(66.8\%,\ 3.3\times 10^{-3})$ & (8,2) & $(71.1\%,\ 4.2\times 10^{-3})$\\
2 & (6,2) & $(78.2\%,\ 2.1\times 10^{-2})$ & (7,2) & $(72.8\%,\ 5.4\times 10^{-4})$ & (8,2) & $(79.6\%,\ 1.0\times 10^{-3})$\\
3 & (6,2) & $(79.1\%,\ 9.5\times 10^{-3})$ & (7,2) & $(77.2\%,\ 9.0\times 10^{-4})$ & (8,2) & $(84.3\%,\ 2.1\times 10^{-4})$\\
4 & (6,2) & $(78.7\%,\ 2.4\times 10^{-3})$ & (7,2) & $(80.1\%,\ 5.7\times 10^{-5})$ & (8,2) & $(86.4\%,\ 1.0\times 10^{-4})$\\
5 & (6,2) & $(77.7\%,\ 2.9\times 10^{-5})$ & (7,2) & $(82.0\%,\ 5.8\times 10^{-5})$ & (8,2) & $(87.2\%,\ 9.5\times 10^{-5})$\\
6 & (6,2) & $(76.7\%,\ 5.4\times 10^{-5})$ & (7,2) & $(83.0\%,\ 6.1\times 10^{-5})$ & (8,2) & $(87.5\%,\ 9.5\times 10^{-5})$\\
7 & (6,2) & $(76.0\%,\ 3.8\times 10^{-5})$ & (7,2) & $(83.6\%,\ 6.3\times 10^{-5})$ & (8,2) & $(87.6\%,\ 9.6\times 10^{-5})$\\
8 & (6,2) & $(75.4\%,\ 3.7\times 10^{-5})$ & (7,2) & $(83.9\%,\ 6.4\times 10^{-5})$ & (8,2) & $(87.7\%,\ 9.6\times 10^{-5})$\\
\hline
\end{tabular}
\end{table}

\subsubsection{Auxiliary Sources}

\qquad Although nearly perfect sources can be obtained at room temperature~\cite{DS2021Room}, indicating that auxiliary sources will not be a barrier, implementing the ESPD scheme may not require single-photon sources. The key parameter $P$ represents the detection probability of the state $|1\rangle$ from the auxiliary path with an ideal SPD when the control signal input is non-vacuum, which is a global parameter regardless of the exact number of photons or the presence of unwanted $|0\rangle$. Intuitively, multi-photon sources might even offer practical benefits over single-photon sources by providing more detection opportunities for the required state $|1\rangle$\footnote{Since the architecture incorporates a filter block, any additional $|0\rangle$ components will be filtered out. Therefore, for a non-vacuum input of a controlled module, if the auxiliary multi-photon signal results in more state $|1\rangle$, then a positive report has a higher chance to be produced, which provides benefits since it enlarges $P$, while, if the auxiliary multi-photon signal adds state $|0\rangle$, then they will be filtered out, thus providing no effects. Similarly, for a vacuum input of a controlled module, only $|0\rangle$ is added in auxiliary paths, which will be filtered out and thereby have no effect even if the source is multiple-photon.}. However, a rigorous formulation of this effect, including its impact on the controlled operations, remains an open problem and is left for future investigation.

\subsection{Limitations and Future Work}

Despite the significance of the ESPD framework, some limitations remain and warrant future investigation.

\subsubsection{Theoretical Completeness}

The theoretical framework established in Section \ref{sec:Theory} remains an area for deeper exploration. While the Monte Carlo analysis and numerical simulations presented in Section \ref{sec:Experiments} are sufficient for convergence demonstration and physical implementation guidance, a complete theoretical study of the dynamic system in Eq.\eqref{DynamicSystem} remains of fundamental academic interest. Future work should focus on rigorously deducing the exact convergence conditions and calculating the precise steady-state point from a more theoretical perspective. Given the inherent complexity of this non-linear and time-variant system, and the fact that numerical designs already suffice for practical implementations, such specialized theoretical investigations are deferred to future work.

\subsubsection{Resource Overhead vs. Optional Simplicity}
\label{sec:Overhead}

While the parallelized nature of the ESPD detection stages theoretically preserves efficiency, the recursive architecture introduces a resource overhead that increases exponentially with the number of levels. Specifically, an $L$-level ESPD utilizing $n$ auxiliary signals per level requires a total of $(n+1)^{L}$ base detections. Although significant performance gains are typically achievable with moderate depth ($L=2$ or $3$), this inherent scaling increases the architectural footprint and component count compared to a standalone SPD. For instance, a three-level ESPD scheme employing five controlled modules per level ($n=5$) necessitates $6^{3}=216$ detections (in the absence of techniques such as multiplexing), resulting in hundreds of integrated components. While not posing a fundamental physical barrier (as discussed in Section \ref{sec:Device}), this resource-intensive nature introduces a non-trivial engineering challenge for immediate large-scale deployment.

However, this complexity is compatible with modern large-scale photonic integrated circuits. Recent advances have successfully demonstrated the integration of hundreds to thousands of components on single chips \cite{WP2018Large,WP2018Multidimensional,PD2019Generation,YL2024A}. Notably, very-large-scale integration of quantum photonic components has already been experimentally demonstrated, with over 2,500 integrated on a single silicon-on-insulator chip~\cite{BF2023Very}, confirming that the integration density required for a multi-level ESPD is fully compatible with standard complementary metal-oxide-semiconductor processes. Furthermore, concerns regarding cumulative loss in cascaded logic gates can be mitigated by emerging low-loss platforms. For instance, silicon nitride waveguides have demonstrated propagation losses as low as 0.2 dB/cm in complex programmable processors \cite{TW20198}, and thin-film lithium niobate platforms offer ultra-low losses of approximately 0.06 dB/cm alongside high-speed modulation capabilities \cite{DS2019Ultra}.

In summary, the hardware complexity of the ESPD should be evaluated as a deliberate engineering trade-off against the necessity of cryogenic operation. The SNSPD mandates bulky, energy-intensive, and expensive cryogenic infrastructure (typically operating below 4 K). The ESPD scheme, conversely, shifts the burden from a demanding operating environment to the domain of on-chip integration. By leveraging large-scale photonic platforms, thousands of components can be lithographically fabricated on a single chip, effectively converting high operational costs and environmental constraints into a one-time design and fabrication cost. This trade-off is particularly advantageous for applications where size, weight, and power constraints prohibit the use of cryogenics, such as satellite-based quantum communication payloads, mobile quantum nodes, or research laboratories lacking dedicated low-temperature facilities.

\subsubsection{Optimization Design}

\begin{figure}[htbp]
    \centering
    \begin{subfigure}[htbp]{0.48\columnwidth}
    \includegraphics[width=\linewidth]{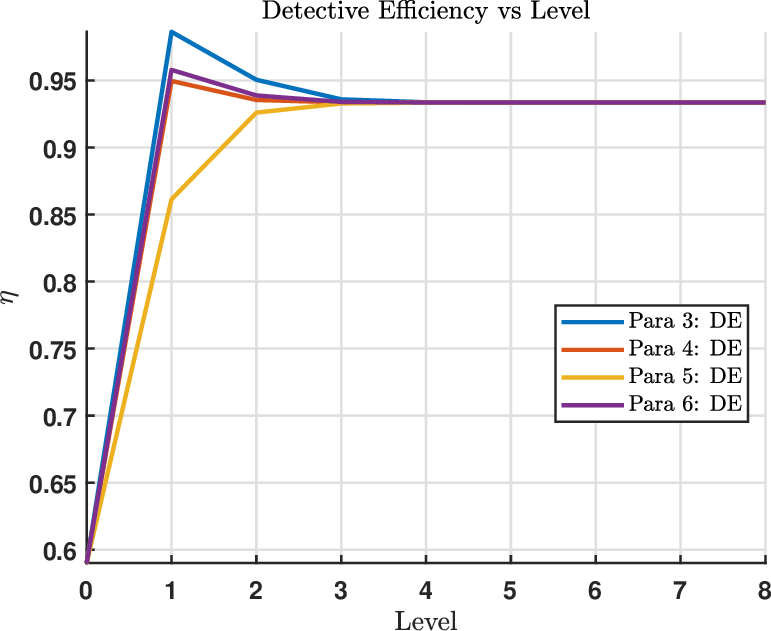}
    \caption{DE $\eta_{s}$ vs.\ level $s$.}
    \end{subfigure}
    \begin{subfigure}[htbp]{0.48\columnwidth}
    \includegraphics[width=\linewidth]{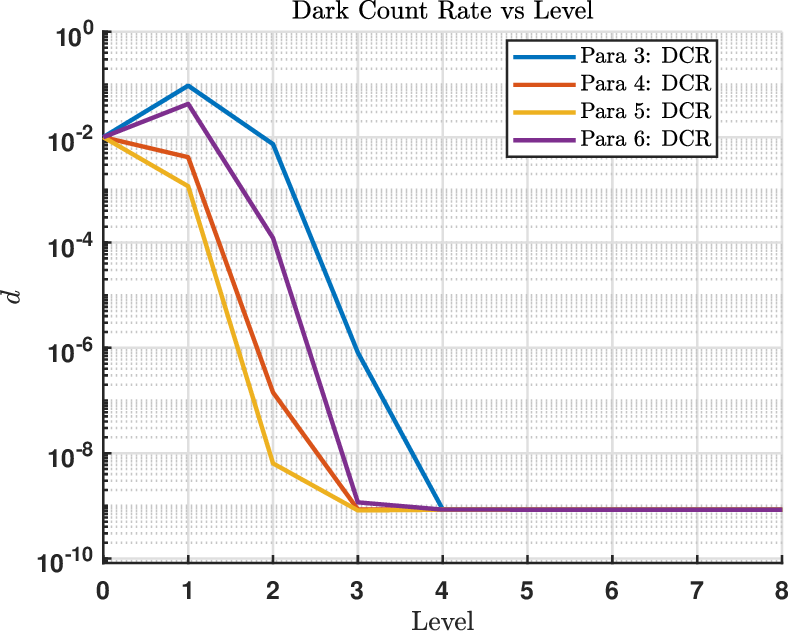}
    \caption{DCR $d_{s}$ vs.\ level $s$.}
    \end{subfigure}
    \caption{\ Performance for parameters $\eta_{0}=59.0\%$, $d_{0}=1.0\times 10^{-2}$, with different values of $n,k$. Para 3 is the same as in Table.\ref{TDE59P97}.}
    \label{FDE59P97n8}
\end{figure}

The inherent exponential resource scaling discussed in Section \ref{sec:Overhead} necessitates optimization for future practical deployments. Specifically, the design of the ESPD requires balancing performance metrics ($DE, DCR$) against implementation complexity ($L, n, k$). Minimizing the iteration depth ($L$) provides significant savings in component count and system complexity due to the exponent dependence. As demonstrated in Section~\ref{sec:Experiments}, the choice of design parameters, namely $n$ (the number of auxiliary signals per level) and $k$ (the decision threshold), plays a critical role: enlarging $k$ reduces DCR exponentially, while improving DE requires a lower $k$ and a larger $n$, with the trade-off being that enlarging $n$ increases detection cost. Furthermore, as shown in Table \ref{TDE59P97n8} and visualized in Fig.\ref{FDE59P97n8}, with a suitably selected $n,k$ on each level, the convergence can be accelerated, and the intermediate-level state can be optimized, which is crucial for minimizing resource consumption.

\begin{table*}[htbp]
\renewcommand\arraystretch{1}
\caption{Performance evolution of the ESPD initialized with a baseline detector from \cite{AW2004Efficient}, with $\eta_{0}=59\%,\ d_{0}=10^{-2}$, under varied $(n_{s},k_{s})$. Settings are the same as in Section \ref{sec:Experiments}}
\centering
\label{TDE59P97n8}
\footnotesize
\begin{tabular}{@{\hspace{0.1mm}}c@{\hspace{1.5mm}}c@{\hspace{1mm}}c@{\hspace{1.5mm}}c@{\hspace{1mm}}c@{\hspace{1.5mm}}c@{\hspace{1mm}}c@{\hspace{0.1mm}}}
\hline
 & \multicolumn{2}{c}{Para 4} & \multicolumn{2}{c}{Para 5} & \multicolumn{2}{c}{Para 6}\\
\hline
Level & $(n,k)$ & (DE\ ($\eta$),\ DCR\ ($d$)) & $(n,k)$ & (DE\ ($\eta$),\ DCR\ ($d$)) & $(n,k)$ & (DE\ ($\eta$),\ DCR\ ($d$))\\
\hline
0 & - & $(59.0\%,\ 1.0\times 10^{-2})$ & - & $(59.0\%,\ 1.0\times 10^{-2})$ & - & $(59.0\%,\ 1.0\times 10^{-2})$\\
1 & (8,2) & $(95.0\%,\ 4.2\times 10^{-3})$ & (4,2) & $(86.1\%,\ 1.2\times 10^{-3})$ & (3,1) & $(95.8\%,\ 4.3\times 10^{-2})$\\
2 & (8,4) & $(93.5\%,\ 1.4\times 10^{-7})$ & (8,4) & $(92.6\%,\ 6.4\times 10^{-9})$ & (6,4) & $(93.9\%,\ 1.2\times 10^{-4})$\\
3 & (8,4) & $(93.4\%,\ 8.5\times 10^{-10})$ & (8,4) & $(93.3\%,\ 8.2\times 10^{-10})$ & (8,4) & $(93.4\%,\ 1.2\times 10^{-9})$\\
4 & (8,4) & $(93.4\%,\ 8.5\times 10^{-10})$ & (8,4) & $(93.3\%,\ 8.4\times 10^{-10})$ & (8,4) & $(93.4\%,\ 8.5\times 10^{-10})$\\
5 & (8,4) & $(93.4\%,\ 8.5\times 10^{-10})$ & (8,4) & $(93.4\%,\ 8.5\times 10^{-10})$ & (8,4) & $(93.4\%,\ 8.5\times 10^{-10})$\\
6 & (8,4) & $(93.4\%,\ 8.5\times 10^{-10})$ & (8,4) & $(93.4\%,\ 8.5\times 10^{-10})$ & (8,4) & $(93.4\%,\ 8.5\times 10^{-10})$\\
7 & (8,4) & $(93.4\%,\ 8.5\times 10^{-10})$ & (8,4) & $(93.4\%,\ 8.5\times 10^{-10})$ & (8,4) & $(93.4\%,\ 8.5\times 10^{-10})$\\
8 & (8,4) & $(93.4\%,\ 8.5\times 10^{-10})$ & (8,4) & $(93.4\%,\ 8.5\times 10^{-10})$ & (8,4) & $(93.4\%,\ 8.5\times 10^{-10})$\\
\hline
\end{tabular}
\end{table*}

The optimization problem also encompasses the optimal circuit topology. To mitigate cumulative error propagation, minimizing the maximal circuit length is essential. As the controlled modules' order can be changed and the auxiliary signals behave similarly to the control signal\footnote{This justifies the dedicated filter block within the paradigm: For the controlled gate be C-NOT, the auxiliary output state is $|1\rangle$ if the input signal is non-vacuum, while is $|\emptyset\rangle$ if the input signal is vacuum, theoretically consistent to the input signal after state-preparation.}, alternative designs could involve leveraging a binary-tree structure and defining the positive report event via a global decision criterion, rather than the current level-by-level $n$-length structure with local (level-wise) criterion used in Section \ref{sec:Design}. However, these architectural changes introduce new theoretical complexity: the general binary-tree structure may result in non-independent distributions that are difficult to analyze, and the global decision criterion selection must carefully account for both device requirements and implementation convenience.

\subsubsection{Physical Integrations and Experiments}

Despite the convincing evidence provided by the numerical simulations in Section~\ref{sec:Experiments}, the physical implementation of the ESPD scheme still requires dedicated experimental validation. Reassuringly, as established in Section \ref{sec:Device}, the scheme relies only on readily available quantum optical components, suggesting no fundamental technological barriers prevent its realization. However, the realization of the large-scale iterative circuit demands significant engineering efforts, particularly concerning the precise integration and synchronization of components. Furthermore, the seamless integration of the ESPD into various quantum optical tasks (such as QKD) needs full investigation. Thus, both the physical realization of the ESPD scheme and the characterization of its performance in applied scenarios are key directions for future work.

\subsubsection{Count Rate and Timing Jitter}

While the ESPD framework significantly enhances DE and suppresses DCR, it generally inherits the timing characteristics of the baseline detectors. Specifically, the system's timing jitter and count rate are primarily determined by the intrinsic temporal response of the baseline SPDs (e.g., the avalanche buildup time and dead time of InGaAs/InP devices). However, the ESPD architecture does not introduce significant additional penalties in principle: the optical-controlled quantum operations function on timescales usually faster than the electronic response of room-temperature SPDs, such that the architectural integration introduces negligible temporal broadening. Furthermore, as the final SPDs operate in parallel, the effective system throughput remains comparable to that of the constituent devices. Nevertheless, future iterations of the ESPD scheme could incorporate techniques such as multiplexing strategies to optimize these temporal metrics as well.

\section{Conclusion}

In summary, we have established the enhanced single-photon detection (ESPD) framework, which reconceptualizes photon detection as an iteratively enhanced quantum-information-processing task that integrates state preparation, controlled operations, projective measurements, and multi-copy decision analysis, rather than as a static device-level operation. Within this system-level theoretical framework, the detection performance is formulated as a non-linear dynamical system, where the effective metrics are progressively optimized through structured quantum processing using exclusively non-superconducting, room-temperature components.

Analytical approximations, Monte Carlo analysis, and numerical simulations, grounded in experimentally reported parameters and commercial photonic specifications, demonstrate that the ESPD framework drives detection performance toward the high-performance basin of attraction, provided the initial conditions lie within a suitable operating range. Consequently, the performance of traditional single-photon detectors (SPDs) can be substantially enhanced: the effective detection efficiency (DE) can be boosted (e.g., from 59\% to 93\%) while the effective dark count rate (DCR) can be suppressed by several orders of magnitude (e.g., from $10^{-2}$ to $10^{-9}$), to the level comparable to state-of-the-art superconducting SPDs, resulting in the significantly expansion of the operational boundaries of quantum communication by relaxing the minimal tolerable channel transmission rate.

Beyond performance enhancement, the ESPD framework establishes a trade-off between resource overhead and environmental constraints, effectively translating the demand for cryogenic cooling during operation into a one-time integration complexity. While large-scale implementation poses engineering challenges in component synchronization and integration density, the presenting theoretical formulations and technological feasibility studies establish a rigorous foundation and provide a roadmap for future experimental realizations.

In essence, this work provides a theoretically grounded framework for high-performance photon detection at room temperature. It demonstrates that intrinsic device-level limitations can be circumvented through system-level architectural design rather than reliance exclusively on material breakthroughs. By establishing the dynamical evolution of the ESPD paradigm, this work not only offers a viable pathway toward next-generation room-temperature photonic devices but also introduces a general methodology for rethinking performance limits in quantum technologies under real-world constraints, potentially influencing a wide range of quantum applications.

\medskip
\textbf{Code Availability} \par 
All data/codes generated or used during this study are included in this published article or can be found at \url{https://github.com/Hao-B-Shu/ESPD}.

\bibliographystyle{unsrt}
\bibliography{QI}

\begin{thebibliography}{10}

\bibitem{GG1999A}
N.~Gisin and B.~Gisin.
\newblock A local hidden variable model of quantum correlation exploiting the
  detection loophole.
\newblock {\em Physics Letters A}, 260(5):323--327, 1999.

\bibitem{MB2023Bounding}
I.~M\'arton, E.~Bene, and T.~V\'ertesi.
\newblock Bounding the detection efficiency threshold in bell tests using
  multiple copies of the maximally entangled two-qubit state carried by a
  single pair of particles.
\newblock {\em Physical Review A}, 107:022205, Feb 2023.

\bibitem{QA2012Maximal}
M.~T. Quintino, M.~Araújo, D.~Cavalcanti, M.~F. Santos, and M.~T. Cunha.
\newblock Maximal violations and efficiency requirements for bell tests with
  photodetection and homodyne measurements.
\newblock {\em Journal of Physics A: Mathematical and Theoretical}, 45, 2012.

\bibitem{VP2010Closing}
T.~V\'ertesi, S.~Pironio, and N.~Brunner.
\newblock Closing the detection loophole in bell experiments using qudits.
\newblock {\em Physical Review Letters}, 104:060401, Feb 2010.

\bibitem{RK2001Experimental}
M.~A. Rowe, D.~Kielpinski, V.~Meyer, C.~A. Sackett, W.~M. Itano, C.~Monroe, and
  D.~J. Wineland.
\newblock Experimental violation of a bell's inequality with efficient
  detection.
\newblock {\em Nature}, 409:791–794, 2001.

\bibitem{M2023Efficiency}
W.~C. Ma.
\newblock Efficiency bounds for testing a multiparticle bell inequality in an
  asymmetric configuration.
\newblock {\em Physical Review A}, 108:052216, Nov 2023.

\bibitem{WR2012Loophole}
B.~Wittmann, S.~Ramelow, F.~Steinlechner, N.~K. Langford, N.~Brunner, H.~M.
  Wiseman, R.~Ursin, and A.~Zeilinger.
\newblock Loophole-free einstein–podolsky–rosen experiment via quantum
  steering.
\newblock {\em New Journal of Physics}, 14:053030, 2012.

\bibitem{E1993Background}
P.~H. Eberhard.
\newblock Background level and counter efficiencies required for a
  loophole-free einstein-podolsky-rosen experiment.
\newblock {\em Physical Review A}, 47:R747--R750, Feb 1993.

\bibitem{CH1974Experimental}
J.~F. Clauser and M.~A. Horne.
\newblock Experimental consequences of objective local theories.
\newblock {\em Physical Review D}, 10:526--535, Jul 1974.

\bibitem{KL2001scheme}
E.~Knill, R.~Laflamme, and G.~A Milburn.
\newblock scheme for efficient quantum computation with linear optics.
\newblock {\em Nature}, 409:46–52, 2001.

\bibitem{DG2025High}
X.~Ding, Y.~P. Guo, M.~C. Xu, R.~Z. Liu, G.~Y. Zou, J.~Y. Zhao, Z.~X. Ge, Q.~H.
  Zhang, H.~L. Liu, L.~J. Wang, M.~C. Chen, H.~Wang, Y.~M. He, Y.~H. Huo, C.~Y.
  Lu, and J.~W. Pan.
\newblock High-efficiency single-photon source above the loss-tolerant
  threshold for efficient linear optical quantum computing.
\newblock {\em Nature Photonics}, 19:387–391, 2025.

\bibitem{SR2005Thresholds}
M.~Silva, M.~R\"otteler, and C.~Zalka.
\newblock Thresholds for linear optics quantum computing with photon loss at
  the detectors.
\newblock {\em Physical Review A}, 72:032307, Sep 2005.

\bibitem{RM2013Highly}
S.~Ramelow, A.~Mech, M.~Giustina, S.~Gr\"{o}blacher, W.~Wieczorek, J.~Beyer,
  A.~Lita, B.~Calkins, T.~Gerrits, S.~W. Nam, A.~Zeilinger, and R.~Ursin.
\newblock Highly efficient heralding of entangled single photons.
\newblock {\em Optics Express}, 21(6):6707--6717, Mar 2013.

\bibitem{NA2015Ultra}
L.~A. Ngah, O.~Alibart, L.~Labonté, V.~D'Auria, and S.~Tanzilli.
\newblock Ultra-fast heralded single photon source based on telecom technology.
\newblock {\em Laser $\&$ Photonics Reviews}, 9, 2015.

\bibitem{DM2022Improved}
S.~I. Davis, A.~Mueller, R.~Valivarthi, N.~Lauk, L.~Narvaez, B.~Korzh, A.~D.
  Beyer, O.~Cerri, M.~Colangelo, K.~K. Berggren, M.~D. Shaw, S.~Xie,
  N.~Sinclair, and M.~Spiropulu.
\newblock Improved heralded single-photon source with a photon-number-resolving
  superconducting nanowire detector.
\newblock {\em Physical Review Applied}, 18:064007, Dec 2022.

\bibitem{SC2023High}
L.~Stasi, P.~Caspar, T.~Brydges, H.~Zbinden, F.~Bussieres, and R.~Thew.
\newblock High-efficiency photon-number-resolving detector for improving
  heralded single-photon sources.
\newblock {\em Quantum Science and Technology}, 8(4):045006, 2023.

\bibitem{BB1984Quantum}
C.~H. Bennett and G.~Brassard.
\newblock Quantum cryptography: Public key distribution and coin tossing.
\newblock In {\em In Proceedings of IEEE International Conference on
  Computers}, 1984.

\bibitem{GL2004Security}
D.~Gottesman, H.~K. Lo, N.~Lütkenhaus, and J.~Preskill.
\newblock Security of quantum key distribution with imperfect devices.
\newblock {\em Quantum Information and Computation}, 4:325--360, 09 2004.

\bibitem{LY2018Overcoming}
M.~Lucamarini, Z.~L. Yuan, J.~F. Dynes, and A.~J. Shields.
\newblock Overcoming the rate–distance limit of quantum key distribution
  without quantum repeaters.
\newblock {\em Nature}, 557(7705):400–403, May 2018.

\bibitem{WH2019Beating}
S.~Wang, D.~Y. He, Z.~Q. Yin, F.~Y. Lu, C.~H. Cui, W.~Chen, Z.~Zhou, G.~C. Guo,
  and Z.~F. Han.
\newblock Beating the fundamental rate-distance limit in a proof-of-principle
  quantum key distribution system.
\newblock {\em Physical Review X}, 9:021046, Jun 2019.

\bibitem{ZZ2022Mode}
P.~Zeng, H.~Y. Zhou, W.~J. Wu, and X.~F. Ma.
\newblock Mode-pairing quantum key distribution.
\newblock {\em Nature Communications}, 13:3903, 2022.

\bibitem{KY1986Transmission}
H.~Kanamori, H.~Yokota, G.~Tanaka, M.~Watanabe, Y.~Ishiguro, I.~Yoshida,
  T.~Kakii, S.~Itoh, Y.~Asano, and S.~Tanaka.
\newblock Transmission characteristics and reliability of pure-silica-core
  single-mode fibers.
\newblock {\em Journal of Lightwave Technology}, 4(8):1144--1150, 1986.

\bibitem{NK2002Ultra}
K.~Nagayama, M.~Kakui, M.~Matsui, T.~Saitoh, and Y.~Chigusa.
\newblock Ultra-low-loss (0.1484 db/km) pure silica core fibre and extension of
  transmission distance.
\newblock {\em Electronics Letters}, 38:1168--1169, 2002.

\bibitem{HH2013Record}
M.~Hirano, T.~Haruna, Y.~Tamura, T.~Kawano, S.~Ohnuki, Y.~Yamamoto, Y.~Koyano,
  and T.~Sasaki.
\newblock Record low loss, record high fom optical fiber with manufacturable
  process.
\newblock In {\em Optical Fiber Communication Conference/National Fiber Optic
  Engineers Conference 2013}, number PDP5A.7, 2013.

\bibitem{TS2017Lowest}
Y.~Tamura, H.~Sakuma, K.~Morita, M.~Suzuki, Y.~Yamamoto, K.~Shimada, Y.~Honma,
  K.~Sohma, T.~Fujii, and T.~Hasegawa.
\newblock Lowest-ever 0.1419-db/km loss optical fiber.
\newblock In {\em 2017 Optical Fiber Communications Conference and Exhibition
  (OFC)}, pages 1--3, 2017.

\bibitem{S2023Solve}
H.~Shu.
\newblock Solve single photon detector problems.
\newblock {\em Quantum}, 7:1187, November 2023.

\bibitem{S2025Elimiating}
Hao Shu.
\newblock Empty-signal detection: Proof-of-principle scheme for arbitrarily
  long-distance quantum communication, 2025.

\bibitem{AW2004Efficient}
M.~A. Albota and F.~N.~C. Wong.
\newblock Efficient single-photon counting at 1.55 um by means of frequency
  upconversion.
\newblock {\em Optics Letters}, 29(13):1449--1451, Jul 2004.

\bibitem{JL20171.25}
W.~H. Jiang, J.~H. Liu, Y.~Liu, G.~Jin, J.~Zhang, and J.~W. Pan.
\newblock 1.25ghz sine wave gating ingaas/inp single-photon detector with a
  monolithically integrated readout circuit.
\newblock {\em Optics Letters}, 42(24):5090--5093, Dec 2017.

\bibitem{HL2020Detecting}
P.~Hu, H.~Li, L.~X. You, H.~Q. Wang, Y.~Xiao, J.~Huang, X.~Y. Yang, W.~J.
  Zhang, Z.~Wang, and X.~M. Xie.
\newblock Detecting single infrared photons toward optimal system detection
  efficiency.
\newblock {\em Optics Express}, 28(24):36884--36891, Nov 2020.

\bibitem{CL2021Detecting}
J.~Chang, J.~W.~N. Los, J.~O. Tenorio-Pearl, N.~Noordzij, R.~Gourgues,
  A.~Guardiani, J.~R. Zichi, S.~F. Pereira, H.~P. Urbach, V.~Zwiller, S.~N.
  Dorenbos, and I.~Esmaeil~Zadeh.
\newblock Detecting telecom single photons with 99.5-2.07+0.5\% system
  detection efficiency and high time resolution.
\newblock {\em APL Photonics}, 6(3):036114, 03 2021.

\bibitem{XZ2021Superconducting}
G.~Z. Xu, W.~J. Zhang, L.~X. You, J.~M. Xiong, X.~Q. Sun, H.~Huang, X.~Ou,
  Y.~M. Pan, C.~L. Lv, H.~Li, Z.~Wang, and X.~M. Xie.
\newblock Superconducting microstrip single-photon detector with system
  detection efficiency over 90\% at 1550nm.
\newblock {\em Photonics Research}, 9(6):958--967, Jun 2021.

\bibitem{CK2023High}
I.~Craiciu, B.~Korzh, A.~D. Beyer, A.~Mueller, J.~P. Allmaras, L.~Narv\'{a}ez,
  M.~Spiropulu, B.~Bumble, T.~Lehner, E.~E. Wollman, and M.~D. Shaw.
\newblock High-speed detection of 1550nm single photons with superconducting
  nanowire detectors.
\newblock {\em Optica}, 10(2):183--190, Feb 2023.

\bibitem{CZ2020Sending}
J.~P. Chen, C.~Zhang, Y.~Liu, C.~Jiang, W.~J. Zhang, X.~L. Hu, J.~Y. Guan,
  Z.~W. Yu, H.~Xu, J.~Lin, M.~J. Li, H.~Chen, H.~Li, L.~X. You, Z.~Wang, X.~B.
  Wang, Q.~Zhang, and J.~W. Pan.
\newblock Sending-or-not-sending with independent lasers: Secure twin-field
  quantum key distribution over 509 km.
\newblock {\em Physical Review Letters}, 124:070501, Feb 2020.

\bibitem{ZL2023Experimental}
L.~Zhou, J.~P. Lin, Y.~M. Xie, Y.~S. Lu, Y.~M. Jing, H.~L. Yin, and Z.~L. Yuan.
\newblock Experimental quantum communication overcomes the rate-loss limit
  without global phase tracking.
\newblock {\em Physical Review Letters}, 130:250801, Jun 2023.

\bibitem{LZ2023Experimental}
Y.~Liu, W.~J. Zhang, C.~Jiang, J.~P. Chen, C.~Zhang, W.~X. Pan, D.~Ma, H.~Dong,
  J.~M. Xiong, C.~J. Zhang, H.~Li, R.~C. Wang, J.~Wu, T.~Y. Chen, L.~X. You,
  X.~B. Wang, Q.~Zhang, and J.~W. Pan.
\newblock Experimental twin-field quantum key distribution over 1000 km fiber
  distance.
\newblock {\em Physical Review Letters}, 130:210801, May 2023.

\bibitem{CB2024Single}
I.~Charaev, E.~K. Batson, S.~Cherednichenko, K.~Reidy, V.~Drakinskiy, Y.~Yu,
  S.~Lara-Avila, J.~D. Thomsen, M.~Colangelo, F.~Incalza, K.~Ilin,
  A.~Schilling, and K.~K. Berggren.
\newblock Single-photon detection using large-scale high-temperature mgb2
  sensors at 20 k.
\newblock {\em Nature Communications}, 15:3973, 2024.

\bibitem{KH2021Room}
S.~Krastanov, M.~Heuck, J.~H. Shapiro, P.~Narang, D.~R. Englund, and K.~Jacobs.
\newblock Room-temperature photonic logical qubits via second-order
  nonlinearities.
\newblock {\em Nature Communications}, 12(191), 2021.

\bibitem{HE2023Efficient}
M.~Heuck, D.~R. Englund, and K.~Jacobs.
\newblock Efficient, high-speed two-photon logic gates at room temperature for
  general-purpose quantum information processing, Jan 2023.
\newblock Filed Mar. 3, 2020, Application No. 16/807,662.

\bibitem{SD2025Experimental}
K.~Sengupta, S.~P. Dinesh, K.~M. Shafi, S.~Asokan, and C.~M. Chandrashekar.
\newblock Experimental realization of universal quantum gates and a six-qubit
  entangled state using a photonic quantum walk.
\newblock {\em Physical Review Applied}, 24:024012, Aug 2025.

\bibitem{AD2025A}
S.~Austin, D.~Devulapalli, K.~Hoang, F.~Zhou, K.~Srinivasan, and A.~V.
  Gorshkov.
\newblock A vapor-cavity-qed system for quantum computation and communication,
  2025.

\bibitem{P2025A}
PsiQuantum.
\newblock A manufacturable platform for photonic quantum computing.
\newblock {\em Nature}, 641(876–883), 2025.

\bibitem{HA2014High}
T.~P. Harty, D.~T.~C. Allcock, C.~J. Ballance, L.~Guidoni, H.~A. Janacek, N.~M.
  Linke, D.~N. Stacey, and D.~M. Lucas.
\newblock High-fidelity preparation, gates, memory, and readout of a
  trapped-ion quantum bit.
\newblock {\em Physical Review Letters}, 113(22), 2014.

\bibitem{MD2005Propagation}
M.~Melchiorri, N.~Daldosso, F.~Sbrana, L.~Pavesi, G.~Pucker, C.~Kompocholis,
  P.~Bellutti, and A.~Lui.
\newblock Propagation losses of silicon nitride waveguides in the near-infrared
  range.
\newblock {\em Applied Physics Letters}, 86(12), 2005.

\bibitem{CL2022Ultra}
A.~Chanana, H.~Larocque, R.~Moreira, J.~Carolan, B.~Guha, E.~G. Melo, V.~Anant,
  J.~D. Song, D.~Englund, D.~J. Blumenthal, K.~Srinivasan, and M.~Davanco.
\newblock Ultra-low loss quantum photonic circuits integrated with single
  quantum emitters.
\newblock {\em Nature Communications}, 13:7693, 2022.

\bibitem{FZ2022Transverse}
L.~T. Feng, M.~Zhang, X.~Xiong, D.~Liu, Y.~J. Cheng, F.~M. Jing, X.~Z. Qi,
  Y.~Chen, D.~Y. He, G.~P. Guo, G.~C. Guo, D.~X. Dai, and X.~F. Ren.
\newblock Transverse mode-encoded quantum gate on a silicon photonic chip.
\newblock {\em Physical Review Letters}, 128(6), 2022.

\bibitem{FZ2025Chip}
L.~T. Feng, M.~Zhang, D.~Liu, Y.~J. Cheng, X.~Y. Song, Y.~Y. Ding, D.~X. Dai,
  G.~P. Guo, G.~C. Guo, and X.~F. Ren.
\newblock Chip-to-chip quantum photonic controlled-not gate teleportation.
\newblock {\em Physical Review Letters}, 135:020802, Jul 2025.

\bibitem{LP2010High}
A.~Laing, A.~Peruzzo, A.~Politi, M.~R. Verde, M.~Halder, T.~C. Ralph, M.~G.
  Thompson, and J.~L. O’Brien.
\newblock High-fidelity operation of quantum photonic circuits.
\newblock {\em Applied Physics Letters}, 97(21):211109, 11 2010.

\bibitem{CR2011Integrated}
A.~Crespi, R.~Ramponi, R.~Osellame, L.~Sansoni, I.~Bongioanni, F.~Sciarrino,
  G.~Vallone, and P.~Mataloni.
\newblock Integrated photonic quantum gates for polarization qubits.
\newblock {\em Nature Communications}, 2(566), 2011.

\bibitem{PR2025High}
J.~Piasetzky, A.~Rotem, Y.~Warshavsky, Y.~Drori, K.~Cohen, Y.~Oz, and
  H.~Suchowski.
\newblock High fidelity cnot gates in photonic integrated circuits using
  composite segmented directional couplers, 2025.

\bibitem{NF2025Quantum}
H.~Nakav, T.~Firdoshi, O.~Davidson, B.~C. Das, and O.~Firstenberg.
\newblock Quantum cnot gate with actively synchronized photon pairs, 2025.

\bibitem{GN2025Quantum}
S.~A.~H. Gangaraj and D.~T. Nguyen.
\newblock Quantum photonic gates with two-dimensional random walkers.
\newblock {\em Optics Express}, 33(5):11264--11279, Mar 2025.

\bibitem{DS2021Room}
K.~B. Dideriksen, R.~Schmieg, M.~Zugenmaier, and E.~S. Polzik.
\newblock Room-temperature single-photon source with near-millisecond built-in
  memory.
\newblock {\em Nature Communications}, 12(3699), 2021.

\bibitem{WP2018Large}
J.~Wang, S.~Paesani, Y.~Ding, R.~Santagati, P.~Skrzypczyk, A.~Salavrakos,
  J.~Tura, R.~Augusiak, L.~Mančinska, D.~Bacco, D.~Bonneau, J.~Silverstone,
  Q.~Gong, A.~Acín, K.~Rottwitt, L.~Oxenlowe, J.~O'Brien, A.~Laing, and
  M.~Thompson.
\newblock Large-scale integration of multidimensional quantum photonics
  circuits on silicon.
\newblock In {\em 2018 Conference on Lasers and Electro-Optics (CLEO)}, pages
  1--2, 2018.

\bibitem{WP2018Multidimensional}
J.~W. Wang, S.~Paesani, Y.~H. Ding, R.~Santagati, P.~Skrzypczyk, A.~Salavrakos,
  J.~Tura, R.~Augusiak, L.~Mančinska, D.~Bacco, D.~Bonneau, J.~W. Silverstone,
  Q.~H. Gong, A.~Acín, K.~Rottwitt, L.~K. Oxenløwe, J.~L. O’Brien,
  A.~Laing, and M.~G. Thompson.
\newblock Multidimensional quantum entanglement with large-scale integrated
  optics.
\newblock {\em Science}, 360(6386):285--291, 2018.

\bibitem{PD2019Generation}
S.~Paesani, Y.~H. Ding, R.~Santagati, L.~Chakhmakhchyan, C.~Vigliar,
  K.~Rottwitt, L.~K. Oxenløwe, J.~W. Wang, M.~G. Thompson, and A.~Laing.
\newblock Generation and sampling of quantum states of light in a silicon chip.
\newblock {\em Nature Physics}, 15(9):925–929, 2019.

\bibitem{YL2024A}
S.~Yu, W.~Liu, S.~J. Tao, Z.~P. Li, Y.~T. Wang, Z.~P. Zhong, R.~Patel, Y.~Meng,
  Y.~Z. Yang, Z.~A. Wang, N.~J. Guo, Xiao~D. Z., Z.~Chen, L.~Xu, N.~Zhang,
  X.~Liu, M.~Yang, W.~H. Zhang, Z.~Q. Zhou, and G.~C. Guo.
\newblock A von-neumann-like photonic processor and its application in studying
  quantum signature of chaos.
\newblock {\em Light, science $\&$ applications}, 13:74, 03 2024.

\bibitem{BF2023Very}
J.~M. Bao, Z.~R. Fu, T.~Pramanik, J.~Mao, Y.~L. Chi, Y.~K. Cao, C.~H. Zhai,
  Y.~F. Mao, T.~X. Dai, X.~J. Chen, X.~Y. Jia, L.~S. Zhao, Y.~Zheng, B.~Tang,
  Z.~H. Li, J.~Luo, W.~W. Wang, Y.~Yang, Y.~Y. Peng, and J.~W. Wang.
\newblock Very-large-scale integrated quantum graph photonics.
\newblock {\em Nature Photonics}, 17:1--9, 04 2023.

\bibitem{TW20198}
C.~Taballione, T.~A.~W. Wolterink, J.~Lugani, A.~Eckstein, B.~A. Bell,
  R.~Grootjans, I.~Visscher, D.~Geskus, C.~G.~H. Roeloffzen, J.~J. Renema,
  I.~A. Walmsley, P.~W.~H. Pinkse, and K.~J. Boller.
\newblock 8$\times$8 reconfigurable quantum photonic processor based on silicon
  nitride waveguides.
\newblock {\em Optics Express}, 27(19):26842--26857, Sep 2019.

\bibitem{DS2019Ultra}
B.~Desiatov, A.~Shams-Ansari, M.~Zhang, C.~Wang, and M.~Lon\v{c}ar.
\newblock Ultra-low-loss integrated visible photonics using thin-film lithium
  niobate.
\newblock {\em Optica}, 6(3):380--384, Mar 2019.

\end{thebibliography}

\end{document}